\documentclass[useAMS,usenatbib]{mn2e}
\usepackage{graphics}
\usepackage{epsfig}
\usepackage{latexsym}
\usepackage{amsfonts}
\usepackage{amsmath}
\usepackage{amssymb}

\def\wco{W_\textrm{CO}}

\def\Msol{M_\odot}

\usepackage[english]{babel}
\usepackage{graphicx}
\usepackage{url}
\usepackage{textcomp}

\usepackage[usenames]{color}

\title[Reverberation in the CTA~1 PWN]{
Molecular environment, reverberation, and radiation from the
 pulsar wind nebula in CTA~1}
\author[Mart\'in, Torres, \& Pedaletti]{Jonatan Mart\'in$^{1}$, Diego F. Torres$^{1,2}$ \& Giovanna Pedaletti$^3$\\
$^1$Institute of Space Sciences (IEEC-CSIC), Campus UAB,  c. Can Magrans s/n, 08193 Barcelona, Spain\\
$^2$Instituci\'o Catalana de Recerca i Estudis Avan\c{c}ats (ICREA) Barcelona, Spain\\
$^3$Deutsches Elektronen-Synchrotron (DESY), D-15738 Zeuthen, Germany}

\begin{document}

\date{}

\pagerange{\pageref{firstpage}--\pageref{lastpage}} 
\pubyear{2015}

\maketitle

\label{firstpage}

\begin{abstract}

We estimate the molecular mass around CTA~1 using
data from Planck and the Harvard CO survey. 
We observe that the molecular mass in the vicinity of the complex
is not enough to explain the TeV emission observed by VERITAS, 
even under favorable assumptions for the cosmic-ray acceleration properties of the supernova remnant.
This supports the idea
that the TeV emission comes from the PWN. 
Here, we model the spectrum of the PWN at possible different stages of its evolution, including both
the dynamics of the PWN and the SNR and their interaction via the reverse shock.
We have included in the model the energy lost via radiation by particles and the particles escape when computing the pressure produced by the gas. This
leads to an evolving energy partition, since for the same instantaneous sharing of the injection of energy provided by the rotational
power, the field and the particles are affected differently by radiation and losses.
We present the model, and study in detail how the spectrum of a canonical isolated PWN is affected during compression and re-expansion  
and how this may impact on the CTA~1 case. 
By exploring the phase-space of parameters that lead to radii 
in agreement with those observed, we then analyze different situations that might represent the current stage of the CTA 1 PWN, and discuss caveats and requirements of each one.

\end{abstract}


\section{Introduction}

CTA~1 (G119.5+10.2) was proposed as a supernova remnant (SNR) by \citet{harris1960}.
Based on HI observations, \citet{pineault1993} derived that its distance is 1.4$\pm$0.3 kpc.
In X-rays, CTA~1 was detected for the first
time by {\it ROSAT} \citep{seward1995}. 
The latter authors also reported the presence of a faint compact source
in the central region, RX J0007.0+7302.
\citet{slane1997} detected non-thermal emission coming from it
using {\it ASCA}, indicating the possible presence of a synchrotron nebula. 
{\it Chandra}
observations done by 
\citet{halpern2004} confirmed it, resolving a point-like source embedded in a compact nebula of 3'' in radius, with a jet-like feature.
\citet{lin2012} performed a 105 ks observation with {\it Suzaku} and detected a $\sim$10' extended feature ($\sim 4.1d_{1.4}$ pc),
which was interpreted as the bow shock of the pulsar wind nebula (PWN). 
They fit the spectrum using a power-law with an index of
1.8 suggesting a slow cooling synchrotron scenario.

Based on X-ray thermal emission, \citet{slane1997} suggested CTA 1 has an age of $\sim$20 kyr.
Later, 
\citet{slane2004} improved on this estimate
based on a Sedov expansion model, and proposed that the age is
$\sim$13 kyr.
The latter value is in agreement with the characteristic age
of the pulsar ($\sim 13.9$ kyr).

At high energies, \citet{mattox1996} proposed that the EGRET source 3EG J0010+7309 (which lies in spatial coincidence with
RX J0007.0+7302), was a potential candidate for a radio-quiet $\gamma$-ray pulsar. 
\citet{brazier1998} also pointed out that this
source was pulsar-like, but a search for $\gamma$-ray pulsations using EGRET data failed \citep{ziegler2008}. 
Flux variability and correlation studies with the position of the SNR using EGRET data gave support to this interpretation 
 \citep{romero99,torres03}. 

A radio-quiet
pulsar in CTA~1 was finally discovered during the commissioning phase of the {\it Fermi} satellite \citep{abdo2008}. 
X-ray pulsations
from this source were detected shortly after \citep{caraveo2010,lin2010}. 
The pulsar in CTA~1 has a period of $\sim$316 ms and a spin-down
power of $\sim 4.5 \times 10^{35}$ erg s$^{-1}$ \citep{abdo2012}. 
No radio counterpart to RX J0007.0+7302 was identified, most
likely due to beaming. 
No optical counterpart is known either \citep{mignani2013}.

A few models of the  spectral energy distribution (SED) of the nebula in CTA~1 were presented.
\citet{zhang2009} proposed a model using 
the SNR radio flux as if it was coming entirely from the PWN, an assumption that may be considered as hardly tenable. 
Nevertheless, they
expected that the PWN in CTA~1 could be detected with the {VERITAS} array. 

The TeV detection of a source at the position of the CTA~1 was confirmed by
\citet{aliu2013}, who reported an extended source with dimensions of
$0.3^\circ \times 0.24^\circ$ ($7.3d_{1.4} \times 5.9d_{1.4}$ pc) at 5' from the {\it Fermi} $\gamma$-ray pulsar PSR J0007.0+7303.
A hint for a detection of an extended GeV source in the off-pulse emission of PSR J0007.0+7303  was reported by \citet{abdo2012},
using 2 years of {\it Fermi}-LAT data. However, only upper limits were imposed.
The use of the newest response functions of the {\it Fermi} satellite, together with several years of additional data, apparently does not confirm
this earlier hint and rather verifies the upper limits to the GeV emission of the CTA~1 PWN (Li et al. ({\it Fermi}-LAT Collaboration)
2016, in preparation).

More recently, \citet{aliu2013} produced a 10 kyr model based on the PWN setup described in
\citet{gelfand2009}, using a low ejected mass of 6.1$M_\odot$, a radius of $\sim$7 pc, and a magnetic fraction (see Section 3 below) of 0.2 (implying  a magnetic
field of 6.3 $\mu$G). 
\citet{torres2014} performed a population study of young ($t_{age} \sim 10$ kyr or less) TeV-detected PWNe
using a time-dependent model.
The fit for the PWN in CTA~1 led to an
age of 9 kyr and an ejected mass of 10M$_\odot$, a
radius of 8 pc, and a magnetic fraction of 0.4. 
Such high magnetic fraction, as the one 
obtained by \citet{aliu2013}, contrast with all others magnetic fractions found for young PWNe, which
are in the range 0.01--0.05. The nature of this difference is unclear. 
If the CTA~1 PWN is in, or has passed a reverberation phase, the synchrotron flux could perhaps
be fitted with a lower magnetic fraction. 

In this work, we include a detailed accounting of the dynamical evolution of the PWN  into the radiative model described
by \citet{martin2014} to study the effects of the reverberation phase in
PWNe. 
We then perform a parameter space exploration for the PWN in CTA~1, varying the ejected mass, the age, and the interstellar medium
(ISM) density to see whether it is possible to explain the features of the SED
with a lower magnetic fraction.
We shall also present a detailed study of the molecular environment of the CTA 1 nebula, in order to consider (and discard) 
possible alternative origins of the TeV radiation. We shall particularly analyze the possibility that
the interaction of cosmic-rays with molecular clouds could give rise to the TeV source.
With the latter aim, we shall present  CO data from the Planck satellite and the Harvard survey of \cite{newdame_co}.

\section{The molecular environment in the vicinity of CTA~1}

\begin{figure*}
\begin{center}
\includegraphics[width=0.3 \linewidth]{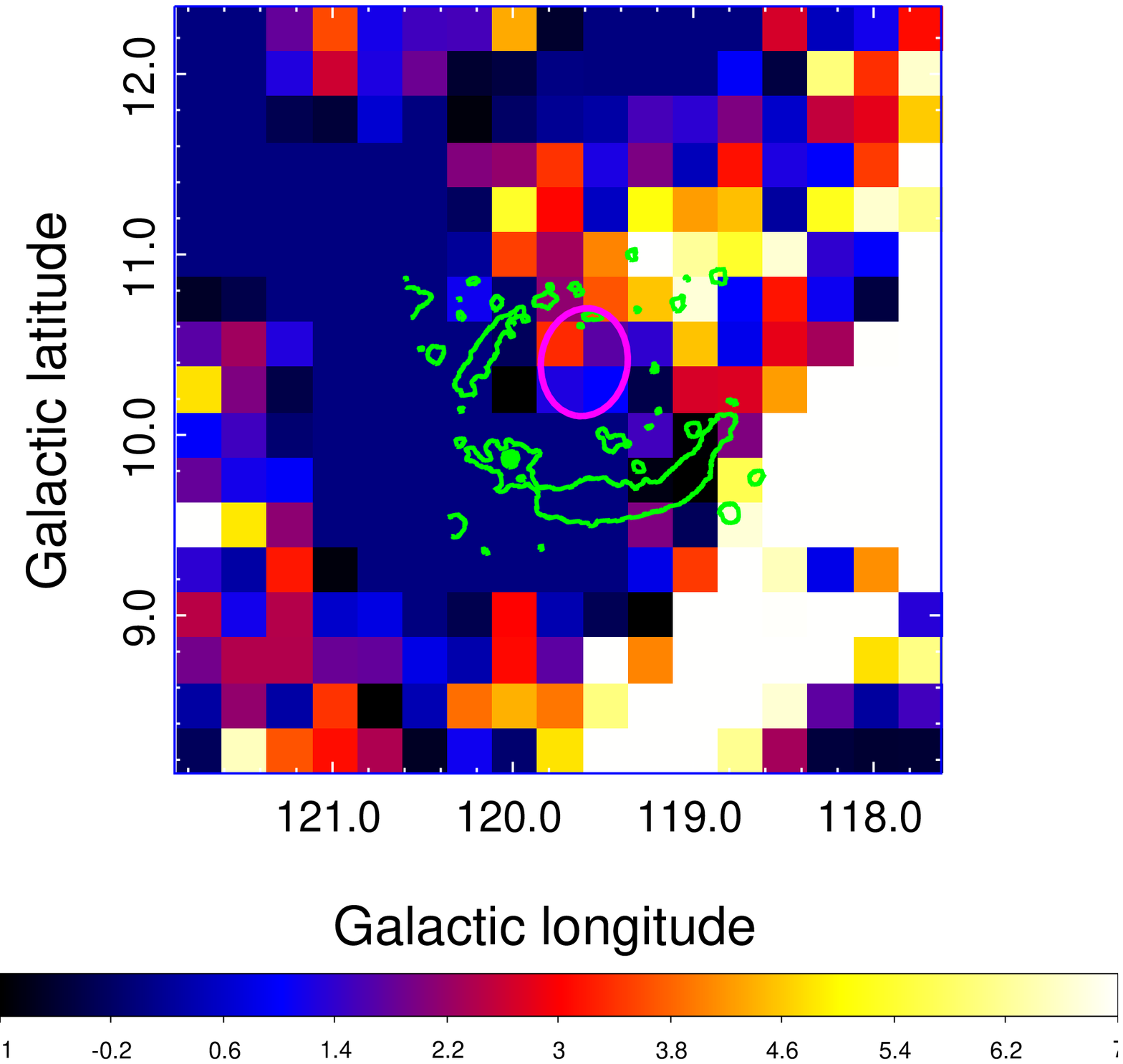}
\includegraphics[width=0.3 \linewidth]{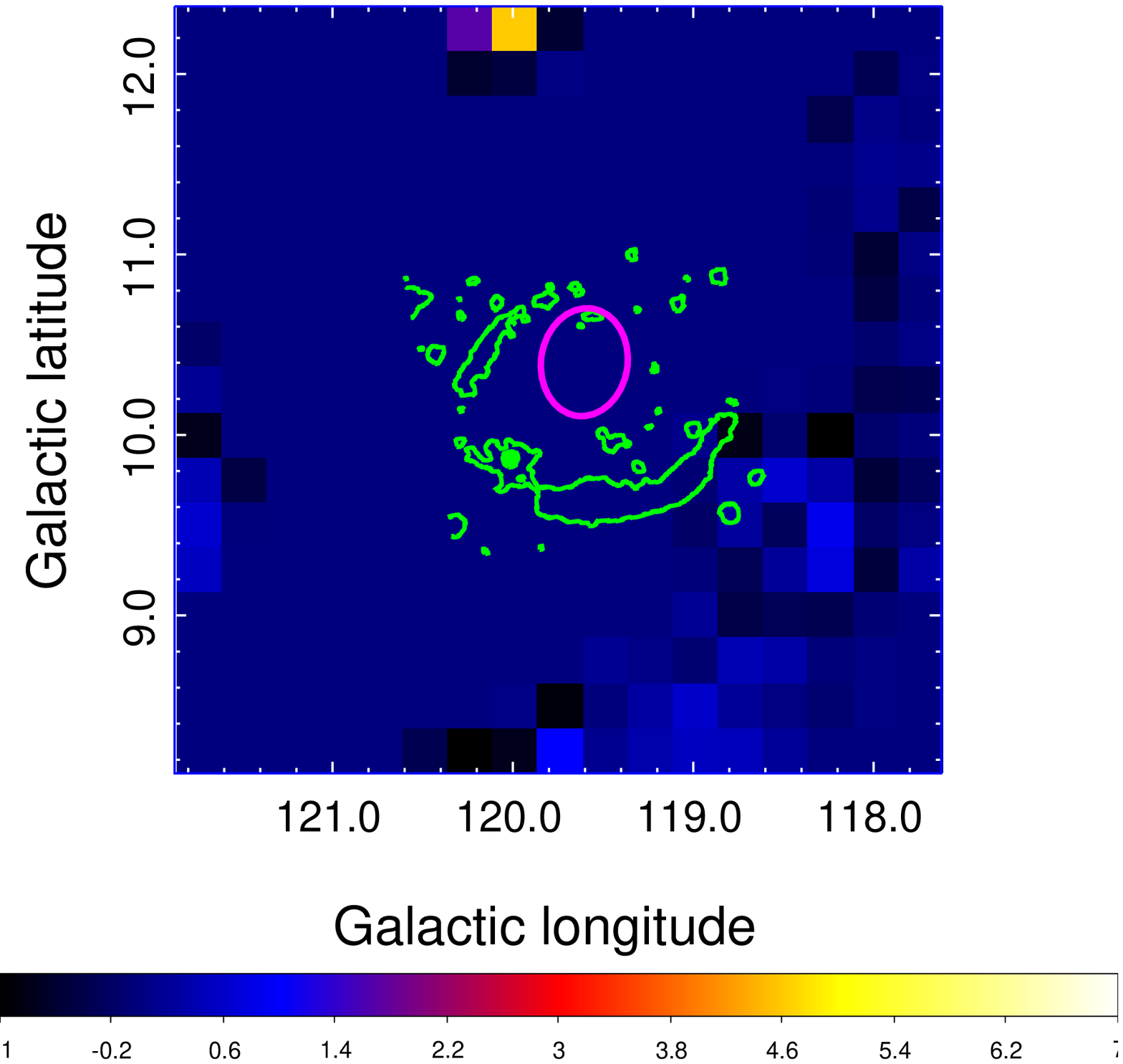}
\includegraphics[width=0.3 \linewidth]{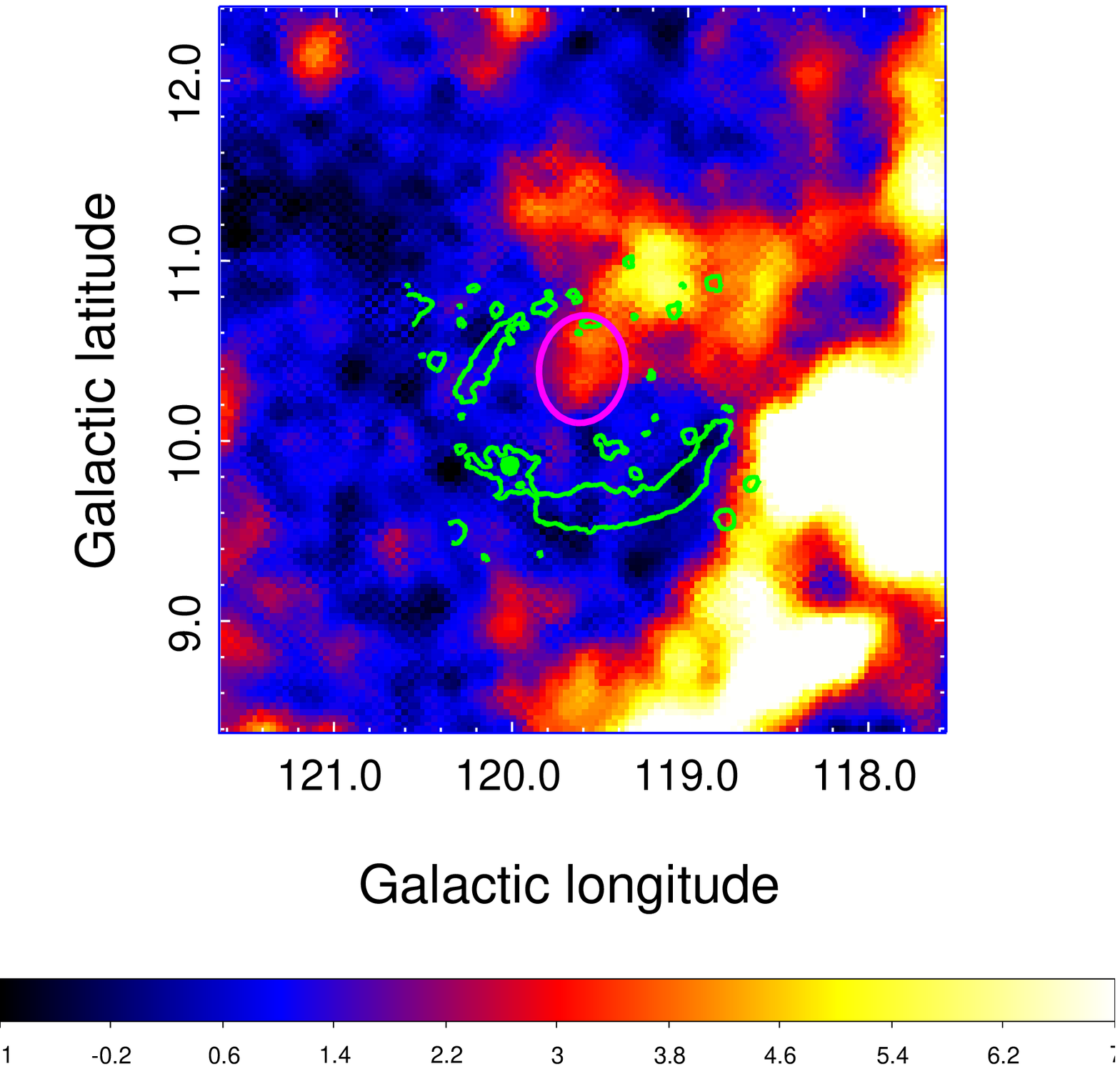}
\includegraphics[width=0.3 \linewidth]{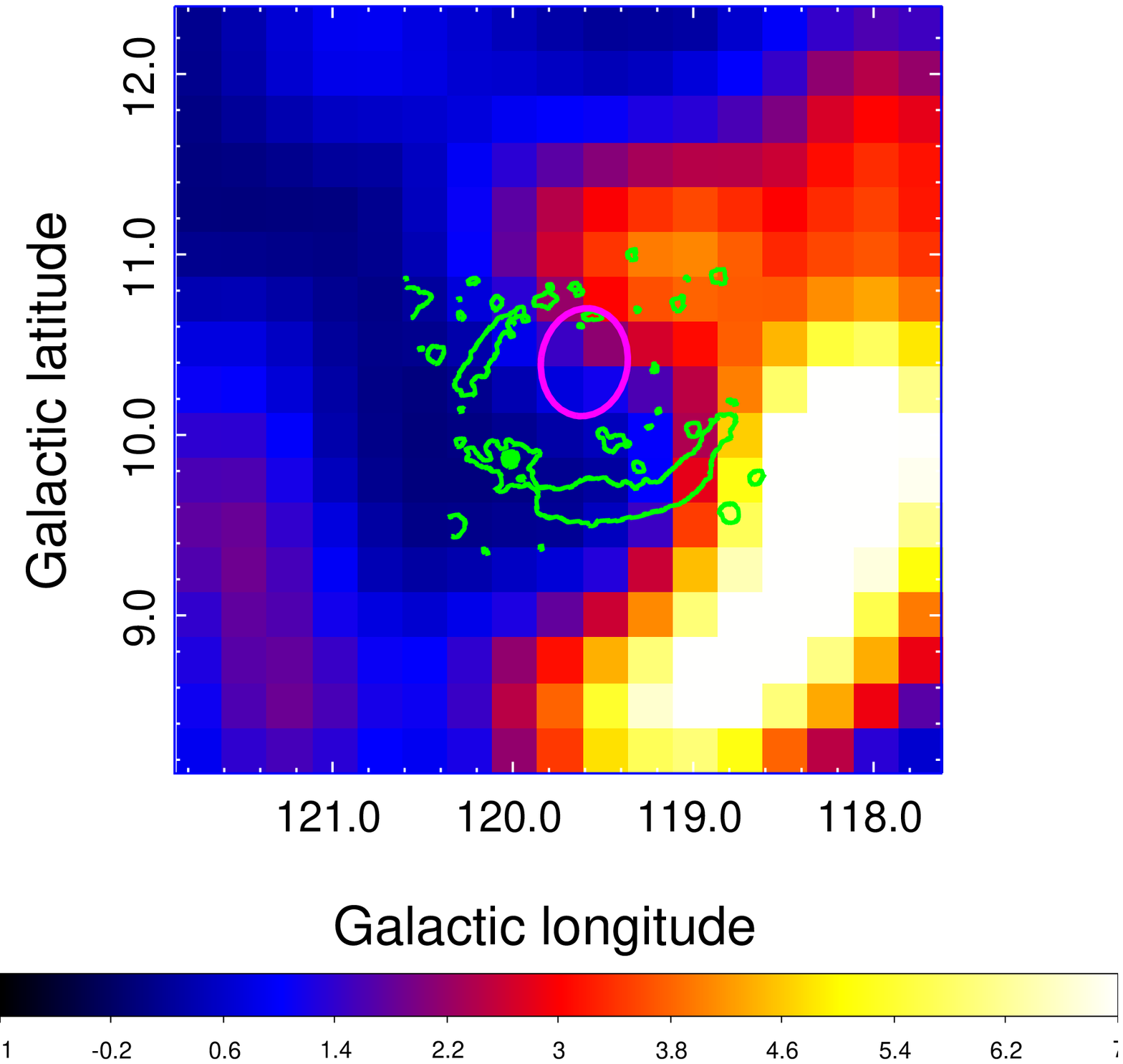}
\includegraphics[width=0.3 \linewidth]{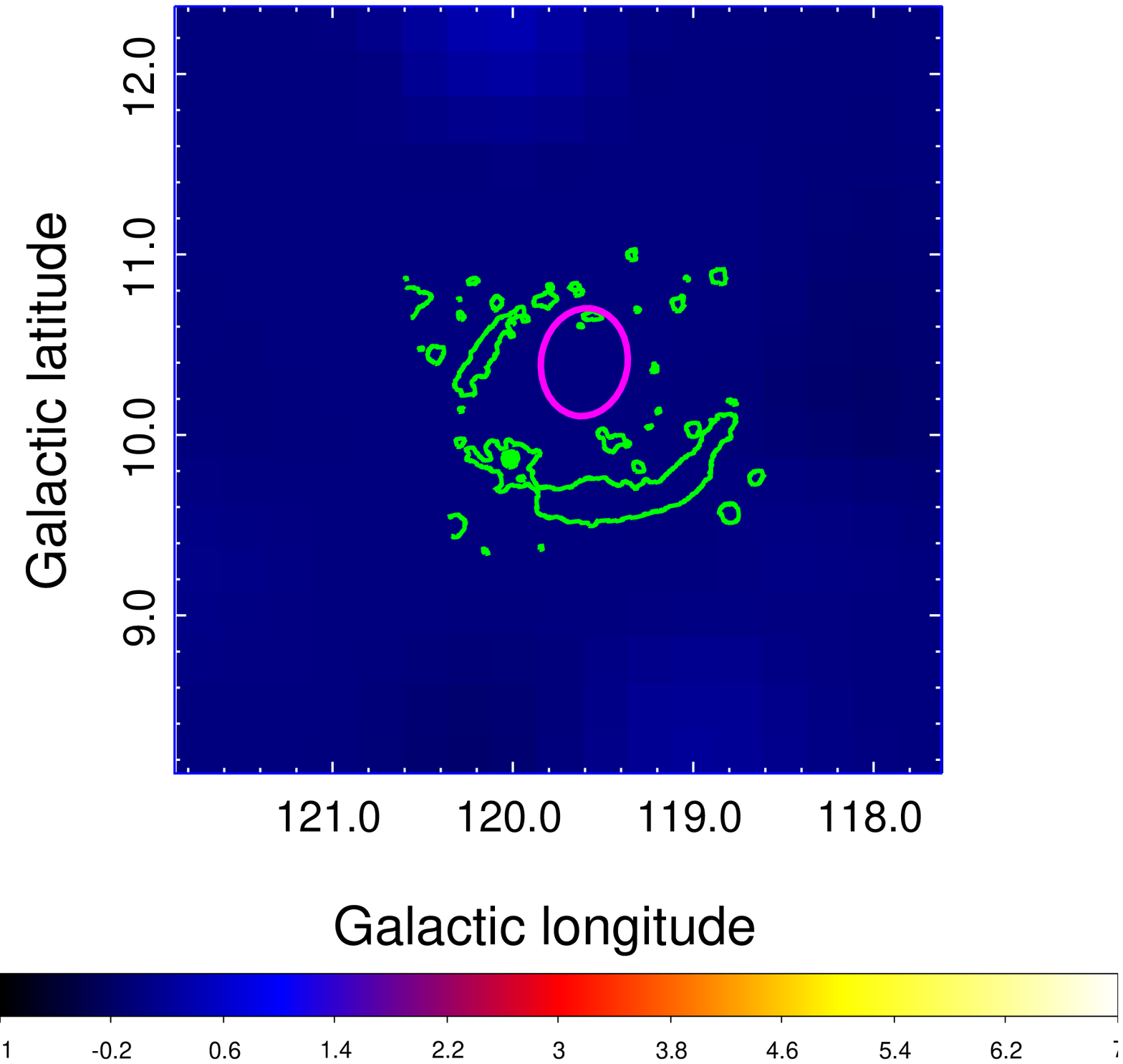}
\includegraphics[width=0.3 \linewidth]{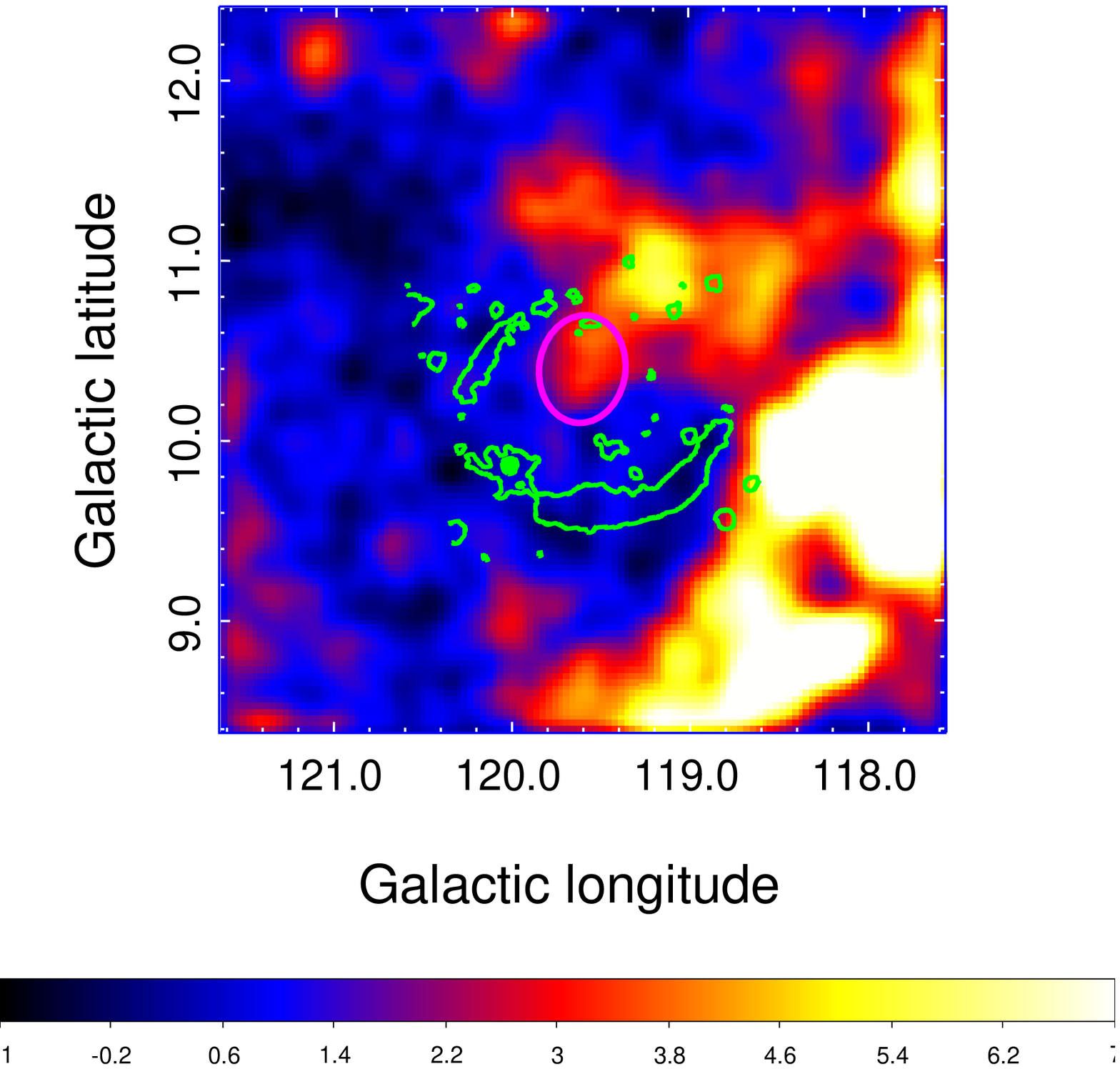}
\caption{$\wco$ of the region surrounding CTA1, in units of (K km/s). Overlaid are the GB6 image countour (4850Mhz) in green and the VERITAS source ellipse in magenta. \textbf{Left:} DHT dataset, integrated along the whole velocity range. \textbf{Center:} DHT dataset, integrated in the velocity range of the HI shell found by \citet{pineault1993}. \textbf{Right:} Planck dataset. Images are on saturated scale to highlight the faint emission in the region of CTA 1.
\textbf{Second row:} same as row above, but smoothed with a 3 pixel Gaussian kernel (1px=15'' for DHT and 1px=2'' for Planck).}
\label{fig:dameco}
\end{center}
\end{figure*}

The molecular content of a region can be estimated through the intensity of the $^\textrm{12}$CO(1$\rightarrow$0) line at 115 GHz \citep[2.6 mm, see e.g.,][]{dame_co}.
A survey of the intensity of the CO line, $\wco$, is provided in \citet{dame_co} and \citet{newdame_co}, hereafter referred to as DHT dataset. 
The data are released\footnote{http://www.cfa.harvard.edu/rtdc/CO/} in cubes of radial velocity, galactic latitude, and longitude. 
The radial velocity connects to the estimation of the distance. 
From $\wco$ one can estimate the mass of the molecular material as follows: 
\begin{equation}\label{eq:massco}
 M_\textrm{CO}=\mu m_\textrm{H} D^2 \Delta\Omega_\textrm{px} X_\textrm{CO} \displaystyle{\sum\limits_\textrm{px}} \wco,
\end{equation}
where $\mu$ is the mean weight, $m_\textrm{H}$ is the mass of the H nucleon, $D$ is the distance, $\Delta\Omega$ is the solid angle.  
The latter is taken out of the summation as it is the same for each pixel (squares of 0.25$^\circ$ per side). 
$X_\mathrm{CO}=N_\mathrm{H2}/\wco$ is a conversion factor defined from the assumption of proportionality of the CO content to the molecular content. 
Hereafter, recognizing that there is some uncertainty in this value, we will use $\mu=2$ and $X_\textrm{CO} = 2.0 \times 10^{20} \textrm{cm}^{-2} (\textrm{K km/s})^{-1} $. 
This is the average galactic value from the review of \citet{xcoreview}. We shall find that our conclusions below do not depend on the value of 
$X_\textrm{CO} $.

Fig. \ref{fig:dameco} presents the data of \citet{newdame_co}, dubbed DHT18 in the archive, cut at the location of CTA~1 and integrated in appropriate velocity ranges. 
We select the maps that were derived with the moment masking technique for noise suppression \citep{dame_mask}. 
The left panel shows the $\wco$ integrated along the entire velocity range.
The moment masking technique is evidently suppressing as noise most of the region. 
The green countours overlaid in Fig. \ref{fig:dameco} are taken from the emission at 4850 MHz (GB6 survey, \citealt{gb6}\footnote{The data was obtained via the \textit{SkyView} Internet archive, http://skyview.gsfc.nasa.gov/current/cgi/titlepage.pl}). 
The region coincident with the VERITAS source \citep{aliu2013} is at the limit of the 
emission recognized as a real signal.
The middle panel of Fig. \ref{fig:dameco} shows the same data set, but clipped in the velocity range $-26<v<-6$, where $v$ is given in units of km/s. This velocity range corresponds to the distance of CTA~1 and of the HI shell seen by Pineault et al. 1993 ($v=-16.9$ km/s). 
However, most of the data in this velocity range is tagged as noise and clipped away. 
Therefore, considering a circle centered at the VERITAS VHE emission centroid ($l$, $b$) = (119.6$^\circ$,10.4$^\circ$), with an extension of 0.25$^\circ$ (recalling that the VERITAS ellipse is 0.3$^\circ$ x 0.24$^\circ$),
we derive a very conservative upper limit on the mass content of the region using the $\wco$ integrated over all the velocity range: 
the total mass amounts to $M_\mathrm{CO} \sim 1000 \Msol$.

We have checked these results using Planck satellite data. We use the Type-2 CO foregorund map of the second release of the Planck satellite data\footnote{http://irsa.ipac.caltech.edu/data/Planck/release\_2/all-sky-maps/}. 
We shown in Fig.  \ref{fig:dameco} the cut on the region, rebinned in pixels of 2'' from an original resolution of 15''. 
The Planck satellite data on CO can only be compared to the DHT dataset when integrated along the whole range of velocity. 
The Planck Collaboration warns that a factor 1.4 in normalization can be expected in their maps with respect to DHT, due to uncertainties in the Planck satellite observations and the calibration in the DHT survey. 
We find only a 15\% difference in the mass estimation from the DHT and Planck data, which given the caveats above and the differences in binning is to be considered acceptable. This difference is in any case irrelevant for the conclusion that follows.

The predicted flux from a molecular cloud embedded in the sea of CR is \cite{aha_passive,gabici_review,gio}:
\begin{equation}\label{eq:passive}
 F(E> E_\gamma) \sim 1 \times 10^{-13}  \kappa E_\gamma^{-1.7} M_\mathrm{5} D ^{-2} \mathrm{cm}^{-2} \mathrm{s}^{-1},
\end{equation}
where $E_\gamma$ is expressed in TeV, the distance $D$ in kpc, the mass $M_\mathrm{5}=10^5 M_\odot$, and $\kappa$ is the enhancement factor of CRs, assumed to be unity for passive clouds (i.e. a case where only the CR background is included) and larger in the presence of a nearby accelerator. 
We are using the PAMELA results by \citet{Adri2011}
for the Galactic cosmic-ray sea at energies $> 200$ GeV. We note that 
the factor $k$ parametrizes our ignorance of the putative source, hence includes both normalization factor and could be considered 
as an enhancement produced by a hardening of the cosmic-ray spectrum too, at least for relevant CR energies of about 10 TeV.
Considering the $M_\mathrm{CO}$ calculated above, one would have $F(E> 1$ TeV$)\sim k \, 5 \times 10^{-16} \mathrm{cm}^{-2} \mathrm{s}^{-1}$, requiring a very large enhancement factor ($k\,\sim1.5\times 10^3$) to reach the VHE flux of $F(E> 1 $ TeV$)=8.5\times 10^{-13}\mathrm{cm}^{-2} \mathrm{s}^{-1}$ as reported in \citet{aliu2013}. 
This number can be compared with  the case of the interacting SNR/molecular cloud complex W28, which also produces GeV and TeV sources, where the value $k$ ranges from 13 to 32 \citep{Aha2008}. Actually, it is of the order of the cosmic-ray enhancement estimated for the central starburst of a whole active galaxy like M82 or NGC 253 \citep{decea,dom2005}.

Thus, for the TeV emission to originate in a molecular cloud illumination (e.g., see  \citet{ic443})
the whole molecular mass along the line of sight should be concentrated right at the neighbourhood of CTA~1, 
and the latter should be acting as a very powerful accelerator. 
This is not supported by data (see middle panels of Fig.  \ref{fig:dameco}) nor by the arguments in the discussion above, and thus
we believe that this possibility is untenable.

\section{PWN Model description}
\label{model}

The evolution of the pair distribution in the PWN, $N$, as a function of the Lorentz factor $\gamma$ at time $t$ is
described by
\begin{equation}
\label{diffloss}
\frac{\partial N(\gamma,t)}{\partial t}=Q(\gamma,t)-\frac{\partial}{\partial \gamma}\left[\dot{\gamma}(\gamma , t)N(\gamma,t) \right]-\frac{N(\gamma,t)}{\tau(\gamma,t)},
\end{equation}
where the term on the left hand side of the equation is the variation in time of the pair distribution. The second and the third
term on the right hand side take into account the energy losses due to synchrotron, inverse Compton (IC), and Bremsstrahlung
interactions, the adiabatic losses, and the escaping particles (we assume Bohm diffusion), respectively. A detailed description of
these terms and the associated formulae can be found in \citet{martin2012,torres2013b}. 

$Q(\gamma,t)$, the injection function, is
to be assumed as a broken power law
\begin{equation}
Q(\gamma,t)=Q_0(t)\left \{
\begin{array}{ll}
\left(\frac{\gamma}{\gamma_b} \right)^{-\alpha_1}  & \text{for }\gamma \le \gamma_b,\\
 \left(\frac{\gamma}{\gamma_b} \right)^{-\alpha_2} & \text{for }\gamma > \gamma_b,
\end{array}  \right .
\end{equation}
and the normalization term $Q_0(t)$ is computed using the spin-down luminosity of the pulsar $L(t)$
\begin{equation}
(1-\eta)L(t)=\int_{\gamma_{min}}^{\gamma_{max}} \gamma m_e c^2 Q(\gamma,t) \mathrm{d}\gamma,
\label{eqeta}
\end{equation}
where the parameter $\eta$ is  is usually called  the magnetic fraction of the nebula, i.e. the fraction of the energy injected by the pulsar that powers the magnetic field of the PWN at each moment of time, see e.g., \citet{gelfand2009,bucciantini2011,martin2012} and references therein. $\eta$ is a free parameter in the model.
Eq. (\ref{eqeta}) is actually used to define the parameter $\eta$,
effectively assigning a fraction of the total rotational power available at each time, 
to particle's energization. 

However, 
$\eta$ is perhaps more properly named as done in Table 1, as the injection sharing parameter, since it describes the {\it instantaneous}
distribution of the spin-down power into the nebula components.  
Since particles radiate along the evolution and are subject to losses and escape,
the total amount of energy contained by them at any moment 
is not simply the summation of the power injected at all times. The actual energy partition (how much energy is stored in the magnetic field
as opposed to how much energy is stored by the particles) will be relatively
enhanced towards the field content in comparison to the case in which such radiation and losses are neglected. 

Studies of the nebula evolution have taken into account particles' radiative losses for determining the spectral energy distribution, but have not, to our knowledge, consider them 
when computing the evolutionary stage. 
Particles' energy and pressure would be smaller in such cases, for a given value of $\eta$. Put otherwise, if we have equipartition at injection at each time ($\eta=0.5$), the nebula would be magnetically dominated in reality; since the particles will loose energy more efficiently than the field.
Simple as it sounds, taking into account this effect produces some complexities and requires a special algorithm to deal with them, described below.

Additionally, referring to the formulae above, $m_e$ and $c$ are the electron mass and the speed of light, respectively. The minimum energy at
injection is a free parameter. If not otherwise stated, we assume it as $\gamma_{min}=1$. The maximum energy at injection is
determined by the most restrictive of two different criteria (see e.g., \citealt{dejager2009}). On the one hand, we consider the synchrotron
limit where the maximum energy is established by the balance between the energy loss by particles due to synchrotron radiation
and the energy gain during acceleration, which yields
\begin{equation}
\label{gsync}
\gamma^{sync}_{max}(t)=\frac{3 m_e c^2}{4 e} \sqrt{\frac{\upi}{e B(t)}},
\end{equation}
where $e$ is the electron charge and $B$ the mean magnetic field of the PWN. On the other hand, demanding confinement of the
particles inside the termination shock of the PWN where particles are accelerated implies
\begin{equation}
\label{ggyro}
\gamma^{gyro}_{max}(t)=\frac{\epsilon e \kappa}{m_e c^2} \sqrt{\frac{\eta L(t)}{c}},
\end{equation}
being $\kappa$ the magnetic compression ratio (we assume it as 3) and $\epsilon$ is the containment fraction ($\epsilon < 1$),
which is the ratio between the Larmor radius of the pairs and the termination shock radius. 

The spin-down luminosity of the pulsar evolves in time as
\begin{equation}
\label{spindown}
L(t)=L_0 \left(1+\frac{t}{\tau_0} \right)^{-\frac{n+1}{n-1}},
\end{equation}
being $n$ the braking index of the pulsar and $\tau_0$ the initial spin-down age defined as
\begin{equation}
\label{tau0}
\tau_0=\frac{2 \tau_c}{n-1}-t,
\end{equation}
where $\tau_c=P/2\dot{P}$ is the characteristic age of the pulsar.

The magnetic field evolution is calculated as in \citet{torres2013a}.
The variation of the magnetic energy in the PWN is balanced by the fraction of the rotational energy that powers the magnetic field
$\eta L(t)$ and the adiabatic losses due to the expansion of the PWN. This is described by the equation
\begin{equation}
\frac{d W_B(t)}{dt}=\eta L(t)-\frac{W_B(t)}{R(t)} \frac{d R(t)}{dt},
\end{equation}
being $W_B=B^2 R^3/6$ the total magnetic energy. This equation can be solved analytically resulting the following formulae
\begin{equation}
\label{bfield}
B(t)=\frac{1}{R^2(t)}\sqrt{6 \eta \int_0^t L(t') R(t') \mathrm{d}t'}
\end{equation}
Here again, take into account that in the equations above the instantaneous sharing $\eta$ is fixed by Eq. (\ref{eqeta}), and is not varying along the PWN evolution.

The radius of the PWN plays an important role in the evolution of the energy losses and the spectrum.
In order to take into account the effects of the reverberation of the PWN after interacting with the reverse shock produced by the ejecta,
we model as well the evolution of the SNR. 
Specifically, in order to compute the radius of the PWN during the free expansion phase and
compression, we solve the equations given by \citet{chevalier2005}, using a similar prescription as in \citet{gelfand2009}. Specifically, we solve
the following system of equations
\begin{eqnarray}
\label{radevol}
&&\frac{d R(t)}{dt}=v(t),\\
&&M(t)\frac{d v(t)}{dt}=4 \upi R^2(t) \left[P(t)-P_{ej}(R,t) \right],\\
&&\frac{d M(t)}{dt}=4 \upi R^2(t) \rho_{ej}(R,t) (v(t)-v_{ej}(R,t)) \label{eqmass}.
\end{eqnarray}
\begin{figure*}
\begin{center}
\includegraphics[width=0.45\textwidth]{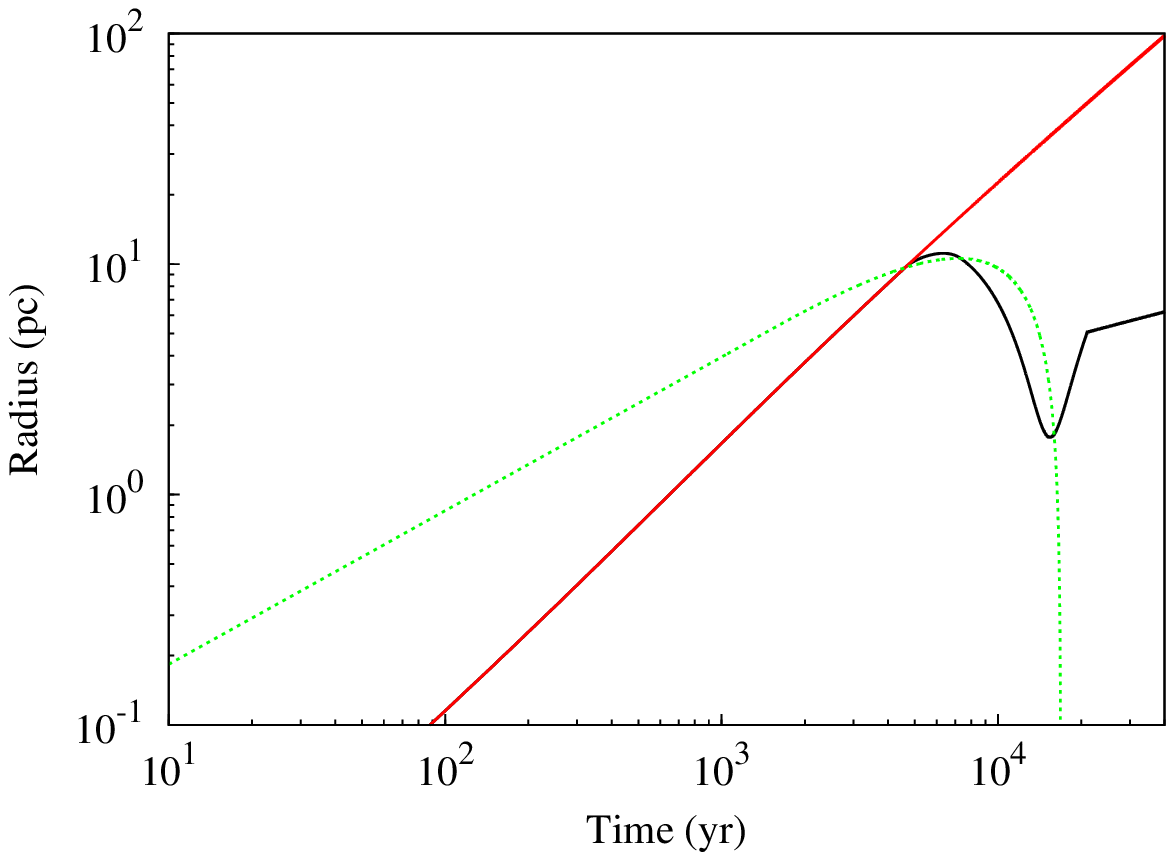}
\includegraphics[width=0.45\textwidth]{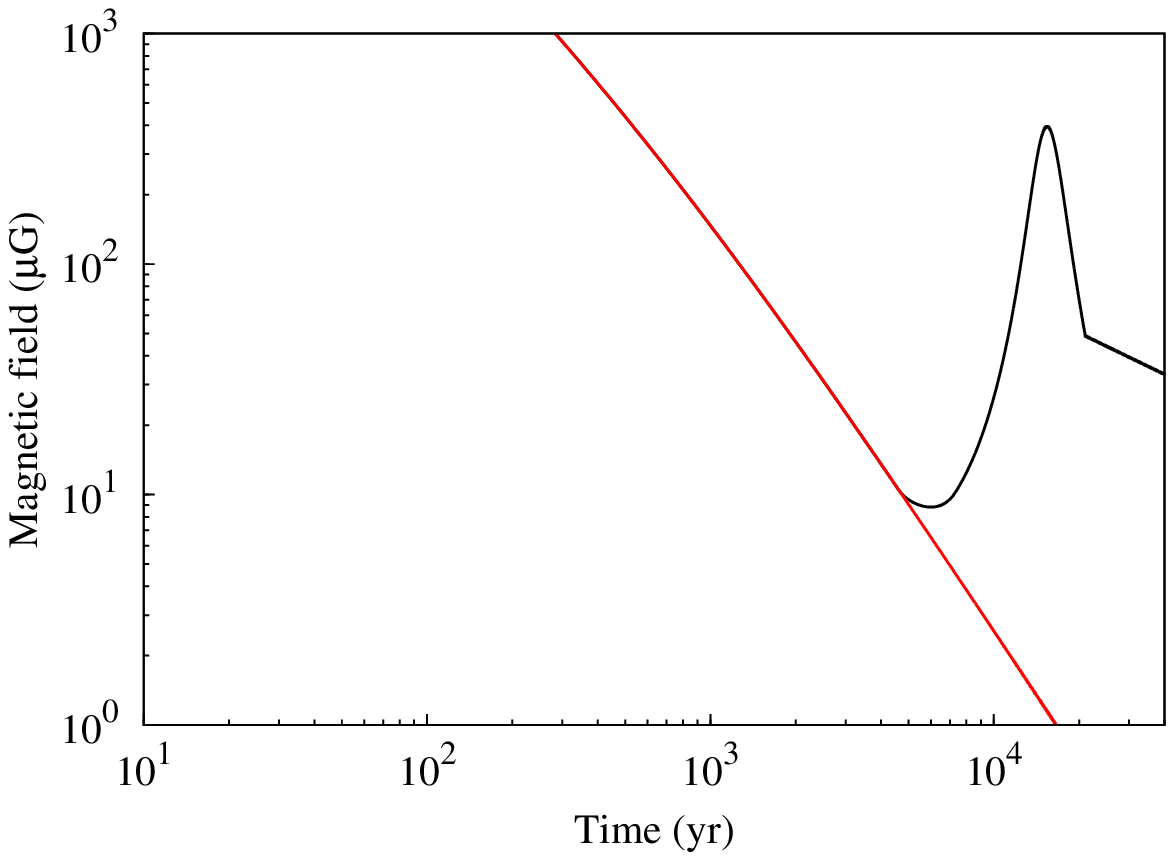}
\end{center}
\caption{Comparison between the radius (left) and the magnetic field (right) for the PWN represented by  the parameters shown
in Table \ref{simulation}. The solid black and red lines show the case where we consider reverberation and only free expansion,
respectively. The dashed green line shows the reverse shock trajectory.}
\label{rad_comp}
\end{figure*}
We note that Eq. (\ref{eqmass}) only applies if $v_{ej}(R,t)<v(t)$. Otherwise, $dM(t)/dt=0$. In the latter set of equations, $v$, $R$,
$M$ and $P$ are the velocity, radius, mass and pressure of the PWN shell. $v_{ej}$, $\rho_{ej}$ and $P_{ej}$ correspond to the
values of the velocity, density, and pressure of the SNR ejecta at the position of the PWN shell. $\gamma_{pwn}$ is the adiabatic
coefficient of the PWN relativistic gas, which is fixed as 4/3. 

We consider now the pressure of the PWN.  
In order to take into account the radiative (as well as other) losses by the particles inside
the PWN in the computation of the pressure $P(t)$, we solve Eq. (\ref{diffloss}) in each time-step and compute the total
energy contained in pairs, $E_p$, by doing the integral,
\begin{equation}
\label{epar}
E_p(t)=\int_{\gamma_{min}}^{\gamma_{max}} \gamma m_e c^2 N(\gamma,t) \mathrm{d}\gamma.
\end{equation}
Since Eq. (\ref{diffloss}) contains a full accounting of all losses when determining $N(\gamma,t)$,  the equation above
is accurately representing the particle's energy share as a function of time.
Then, the pressure contributed by particles is
\begin{equation}
\label{ppar}
P_p(t)=\frac{3(\gamma_{pwn}-1)E_p(t)}{4 \upi R(t)^3}.
\end{equation}
The magnetic field energy also contributes to the total pressure. The magnetic pressure is given by
\begin{equation}
\label{pmag}
P_B(t)=\frac{B^2(t)}{8 \upi}.
\end{equation}
Finally, the total pressure is obtained by summing both contributions 
\begin{equation}
P(t)=P_p(t)+P_B(t).
\end{equation}

As said above, $v_{ej}$, $\rho_{ej}$ and $P_{ej}$ correspond to the
values of the velocity, density, and pressure of the SNR ejecta at the position of the PWN shell.
These profiles change if the PWN shell is surrounded  by unshocked ejecta (thus the radius
of the PWN is smaller than the radius of the reverse shock of the SNR, $R < R_{rs}$), or by shocked ejecta (where $R_{rs} < R < R_{snr}$,
being $R_{snr}$ the radius of the SNR). The initial profiles for the unshocked medium are assumed, following \citet{truelove1999,blondin2001}, as
\begin{equation}
\label{vej}
v_{ej}(r,t)=\left \{
\begin{array}{ll}
r/t & \text{for }r < R_{snr},\\
0 & \text{for }r > R_{snr},
\end{array}  \right .
\end{equation}
\begin{equation}
\label{rhoej}
\rho_{ej}(r,t)=\left \{
\begin{array}{ll}
A/t^3 & \text{for }r < v_t t,\\
A(v_t/r)^\omega t^{\omega-3} & \text{for } v_t t < r < R_{snr},\\
\rho_{ism} & \text{for } r > R_{snr},
\end{array}  \right .
\end{equation}
\begin{equation}
P_{ej}(r,t)=0,
\end{equation}
where
\begin{equation}
A=\frac{(5\omega-25)E_{sn}}{2 \upi \omega v_t^5},
\end{equation}
\begin{equation}
v_t=\sqrt{\frac{10(\omega-5)E_{sn}}{3(\omega-3)M_{ej}}}.
\end{equation}
The parameters $E_{sn}$ and $M_{ej}$ are the energy of the SN and the total ejected mass during the explosion, respectively. In our model,
for simplicity, we assume $\omega=9$ as in \citet{chevalier1992,blondin2001,gelfand2009} for a type II SN. When the PWN shell is surrounded
by the shocked medium, then we use the prescription given by \citet{bandiera1984} to obtain the $v_{ej}$, $\rho_{ej}$ and $P_{ej}$ profiles.
For the shock trajectories, we use the semianalytic model by \citet{truelove1999} for a non-radiative SNR.

After the compression, the PWN bounces and starts the Sedov phase when its pressure reaches the pressure of the SNR's Sedov solution. At
this time, the evolution of the radius of the PWN follows the relation given in \citet{bucciantini2011}
\begin{equation}
\label{sedov}
R^4(t_{Sedov})P(t_{Sedov})=R^4(t)P(t),
\end{equation}
where $P(t)=\rho_{ism} v_{fs}^2/(\gamma_{snr}+1)$ is the pressure in the SNR forward shock. The term $v_{fs}^2$ is the velocity of the
forward shock given by the Truelove \& McKee equations.

The integration of Eq. (\ref{diffloss}), critical in all this modelling, is done following the subsequent prescription:
\begin{enumerate}
 \item We keep the value of the radius $R_{pwn}(t)$. If this is the first time step, the radius is given by \citep{Swaluw}
\begin{eqnarray}
R(\Delta t)=\left(\frac{6}{15(\gamma_{pwn})}+\frac{289}{240} \right)^{-1/5}  \left(\frac{L_0 t}{E_{SN}} \right)^{1/5} \times \nonumber \\
\left(\frac{10E_{SN}}{3M_{ej}} \right)^{1/2} t.
\end{eqnarray}
 \item We proceed to integrate Eqs. (\ref{radevol} -- \ref{eqmass}) with a fourth order Runge Kutta method.
 \item We compute the new values for the spin-down luminosity and magnetic field using Eqs. (\ref{spindown}) and (\ref{bfield}), respectively.
 \item We compute the maximum energy at injection $\gamma_{max}(t)$ using Eqs. (\ref{gsync}) and (\ref{ggyro}), applying the most restrictive criterion.
 \item We numerically integrate the energy loss, Eq. (\ref{diffloss}), to get the electron spectrum $N(t)$.
 \item We obtain the energy and the pressure done by particles by using Eq. (\ref{epar}) and (\ref{ppar}). In this way, we take into account the energy
 losses of the system in order to compute the pressure.
 \item We obtain the pressure done by the magnetic field by using Eq. (\ref{pmag}) and then obtain the total pressure.
 \item We compute the radius and the velocity of the reverse and forward shock of the SNR using the prescription of \citet{truelove1999}.
 \item If the radius of the reverse shock $R_{rs}$ is greater than the radius of the PWN $R_{pwn}$, we approximate the pressure done by the SNR
ejecta as 0, and the velocity and the density of the ejecta are described by Eqs. (\ref{vej}) and (\ref{rhoej}), respectively. If not, we compute these
profiles using the prescription given in \citet{bandiera1984}.
 \item We recurse and go back to the first step.
\end{enumerate}
After step viii, when the first compression has already happened, we use Eqs. (\ref{radevol}) to compute the radius until we reach the pressure
of the Sedov solution for the SNR. At this moment, the evolution of the radius is described by Eq. (\ref{sedov}).

\section{Spectral impact of reverberation}

\begin{table}
\centering
\scriptsize
\vspace{0.2cm}
\caption{Input parameters used in the generic PWN simulation.}
\label{simulation}
\begin{tabular}{lll}
\hline
Definition & Parameter & Value\\
\hline
\hline
Age & $t_{age}$ (kyr) & 40\\
Initial spin down luminosity & $L_0$ (erg s$^{-1}$) & $3 \times 10^{39}$\\
Braking index & $n$ & 3\\
Initial spin down age & $\tau_0$ (kyr) & 1\\
Distance & $d$ (kpc) & 2\\
PWN adiabatic coefficient & $\gamma_{pwn}$ & 4/3\\
SNR adiabatic coefficient &$\gamma_{snr}$ & 5/3\\
Index of the SNR density power law & $\omega$ & 9\\
ISM density & $\rho_{ism}$ & 0.1\\
Energy of the SN & $E_{sn}$ & $10^{51}$\\
Ejected mass & $M_{ej}$ (M$_{\odot}$) & 10\\
CMB temperature & $T_{cmb}$ (K) & 2.7\\
CMB energy density & $w_{cmb}$ (eV cm$^{-3}$) & 0.26\\
FIR temperature & $T_{fir}$ (K) & 25\\
FIR energy density & $w_{fir}$ (eV cm$^{-3}$) & 0.5\\
NIR temperature & $T_{nir}$ (K) & 3000\\
NIR energy density & $w_{nir}$ (eV cm$^{-3}$) & 1\\
Minimum energy at injection & $\gamma_{min}$ & 1\\
Energy break at injection & $\gamma_b$ & $10^6$\\
Low energy index at injection & $\alpha_l$ & 1.5\\
High energy index at injection & $\alpha_h$ & 2.5\\
Containment factor & $\epsilon$ & 0.3\\
Injection sharing (magnetic fraction) & $\eta$ & 0.03\\
\hline
\hline
\end{tabular}
\end{table}

\begin{figure*}
\begin{center}
\includegraphics[width=0.45\textwidth]{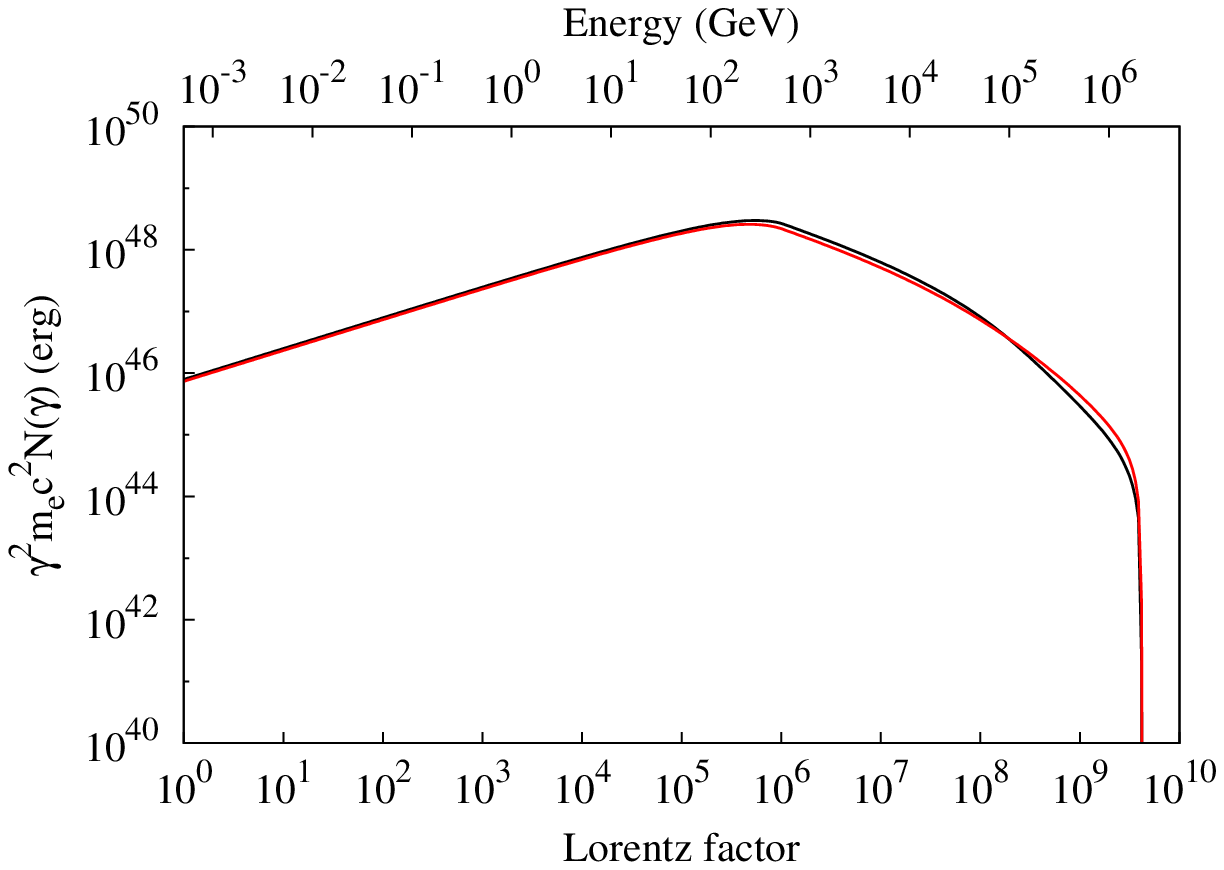}
\includegraphics[width=0.45\textwidth]{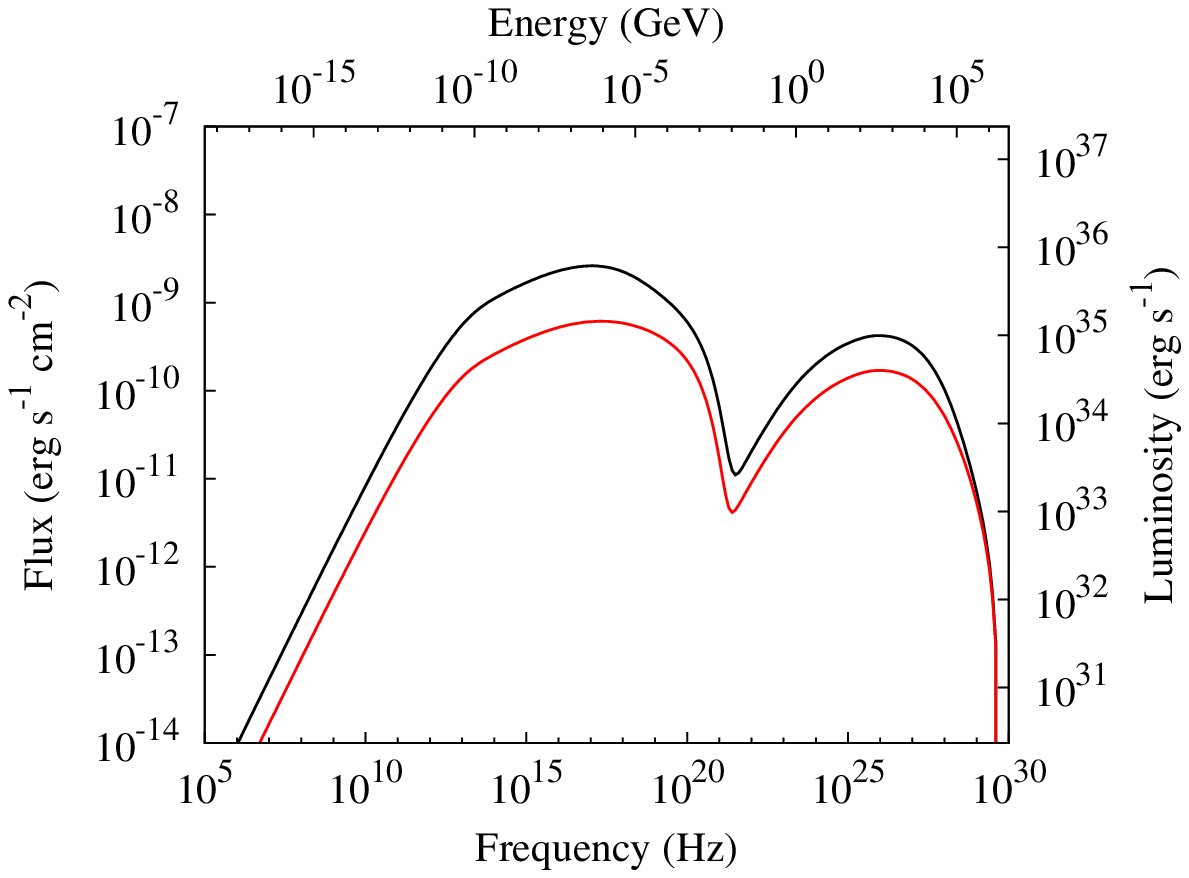}\\
\includegraphics[width=0.45\textwidth]{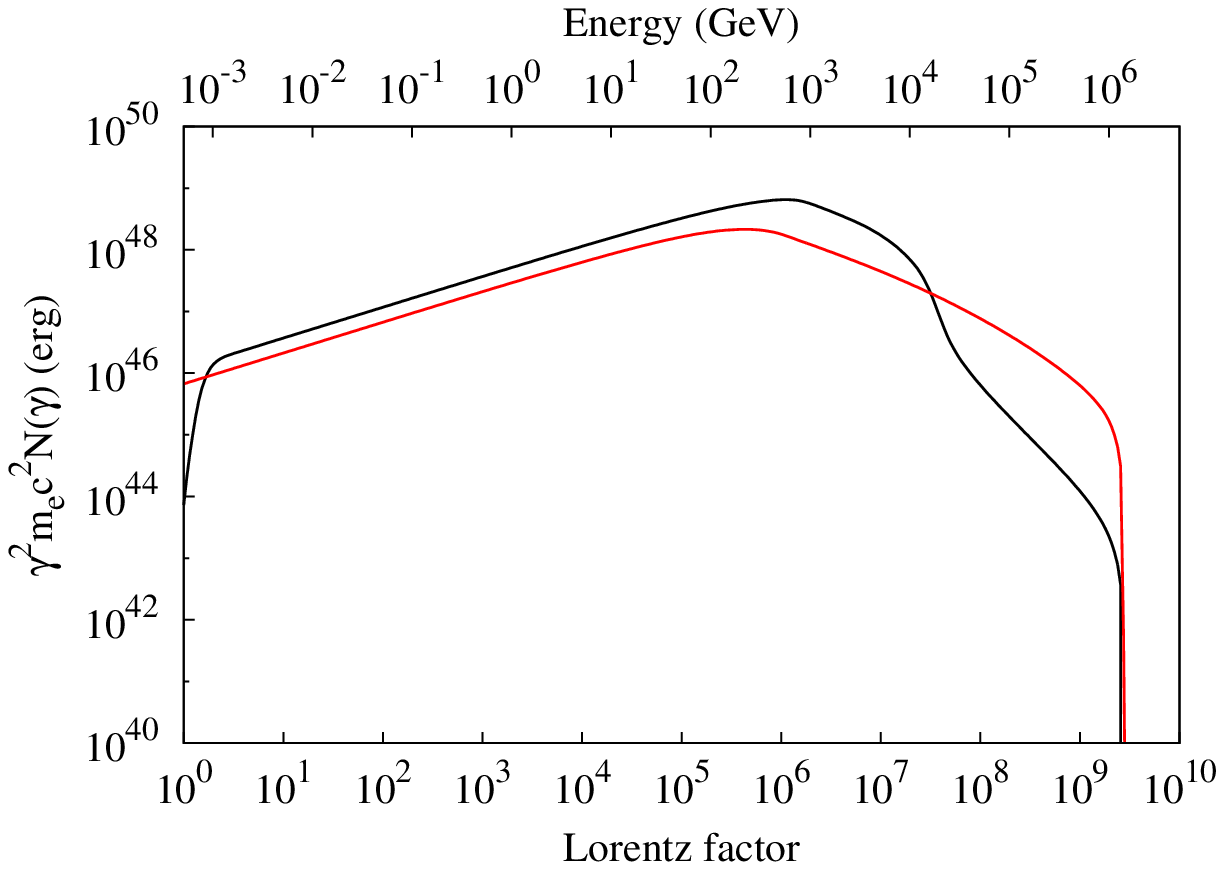}
\includegraphics[width=0.45\textwidth]{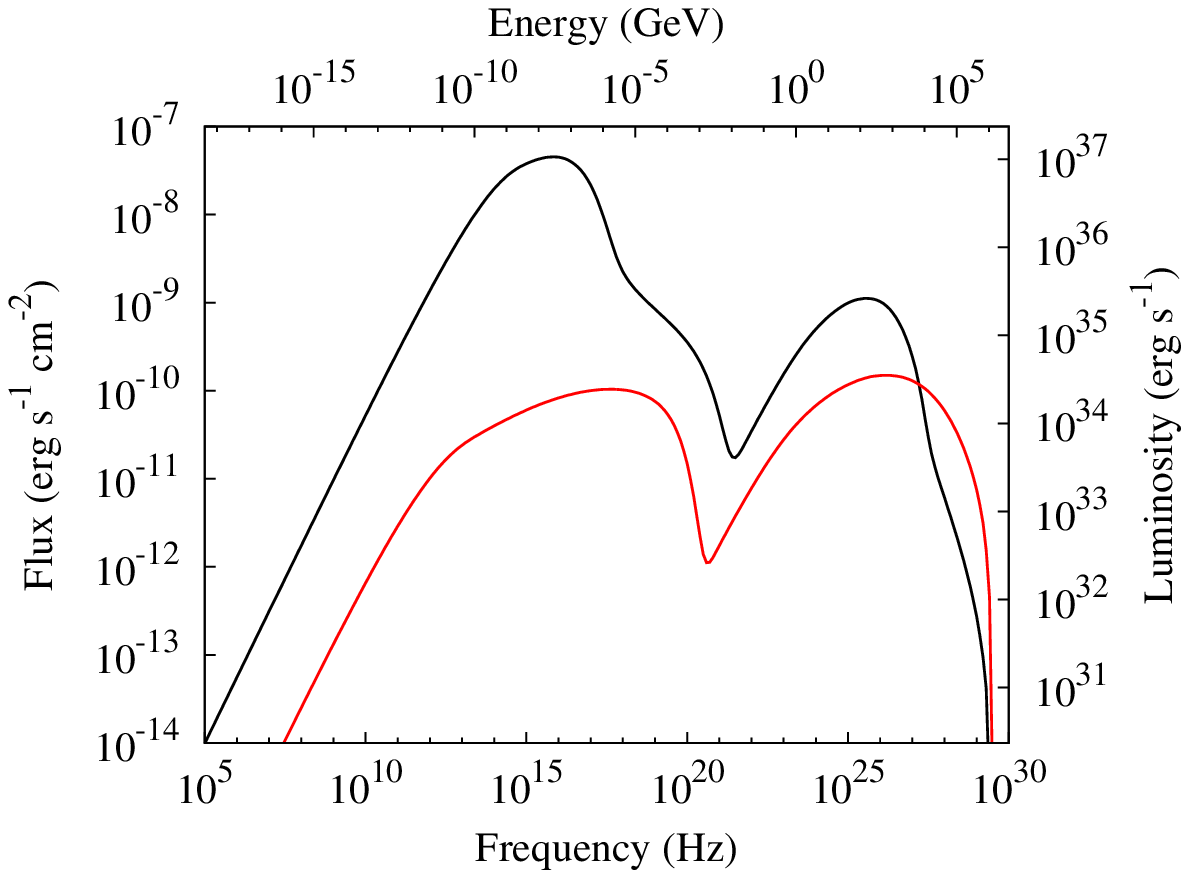}\\
\includegraphics[width=0.45\textwidth]{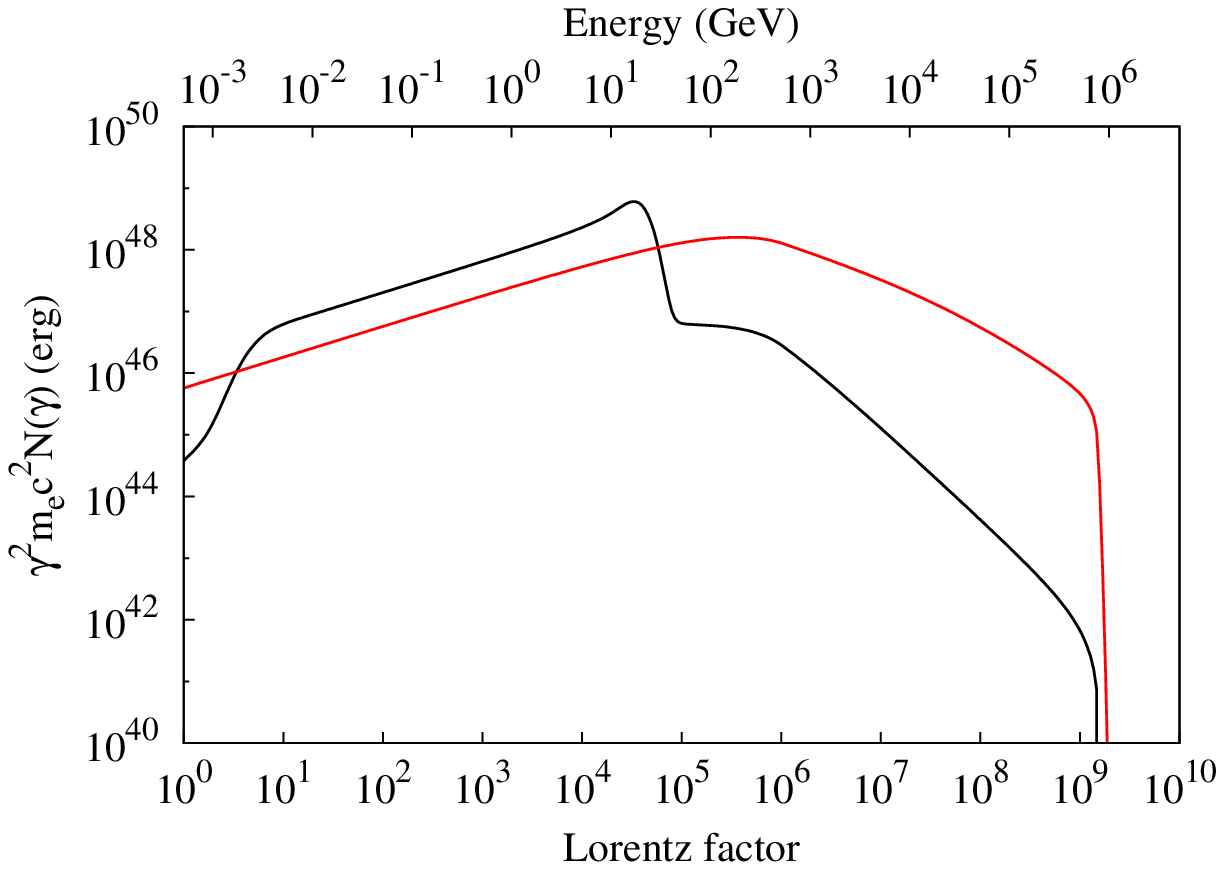}
\includegraphics[width=0.45\textwidth]{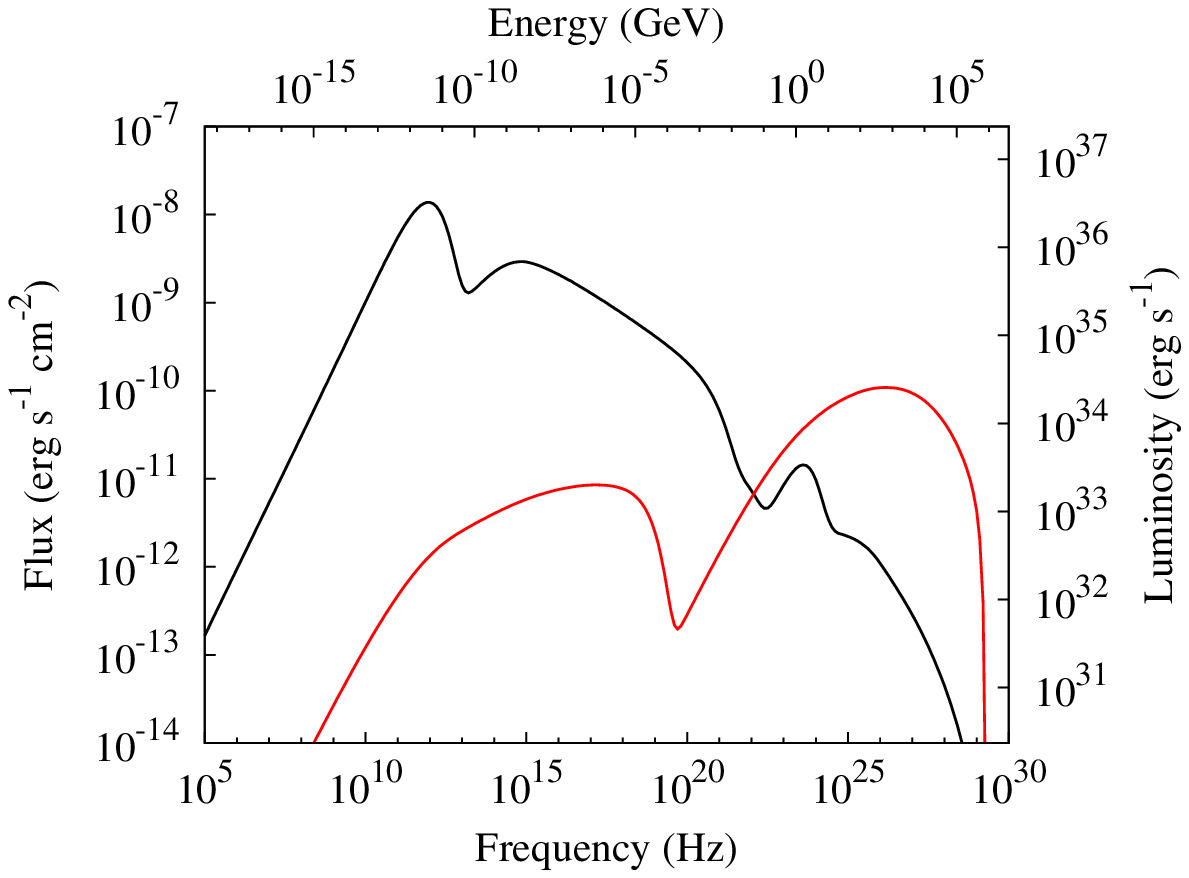}\\
\includegraphics[width=0.45\textwidth]{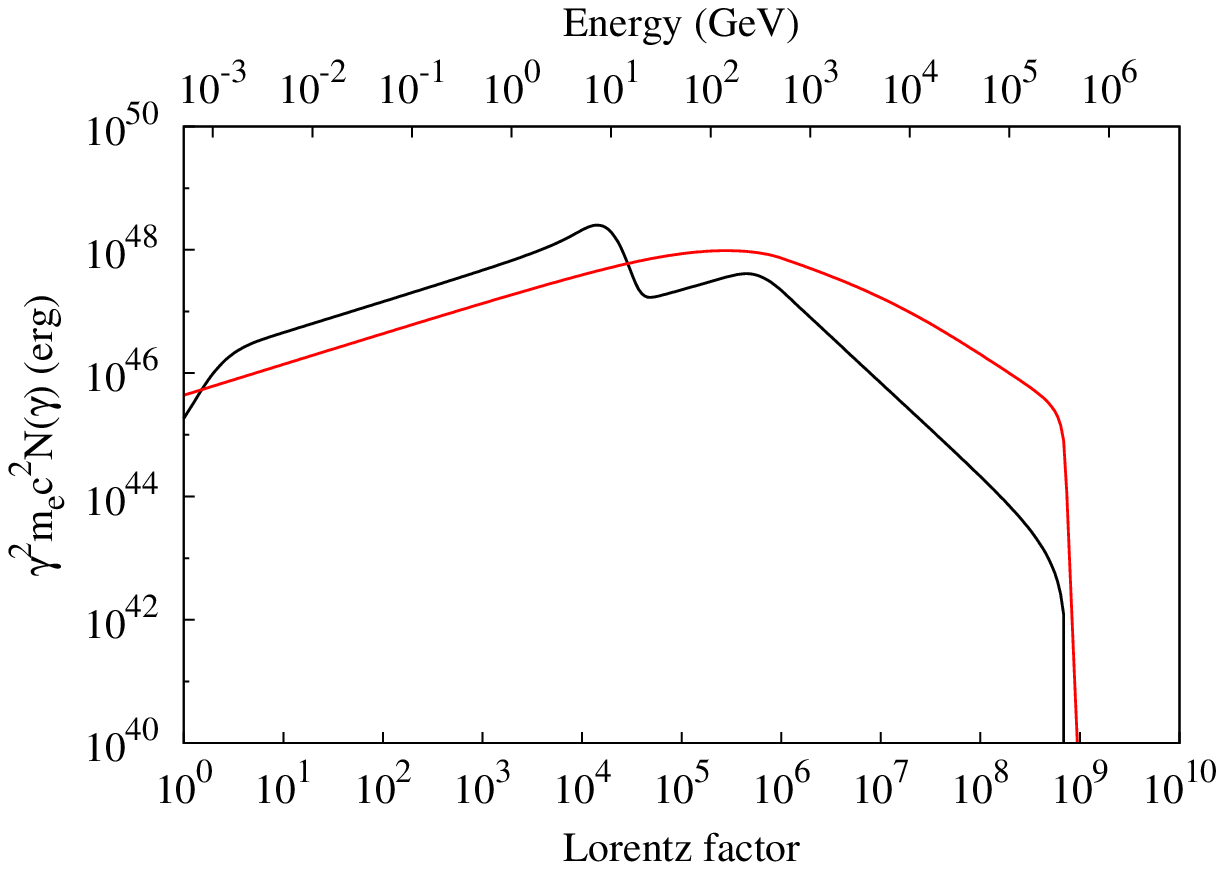}
\includegraphics[width=0.45\textwidth]{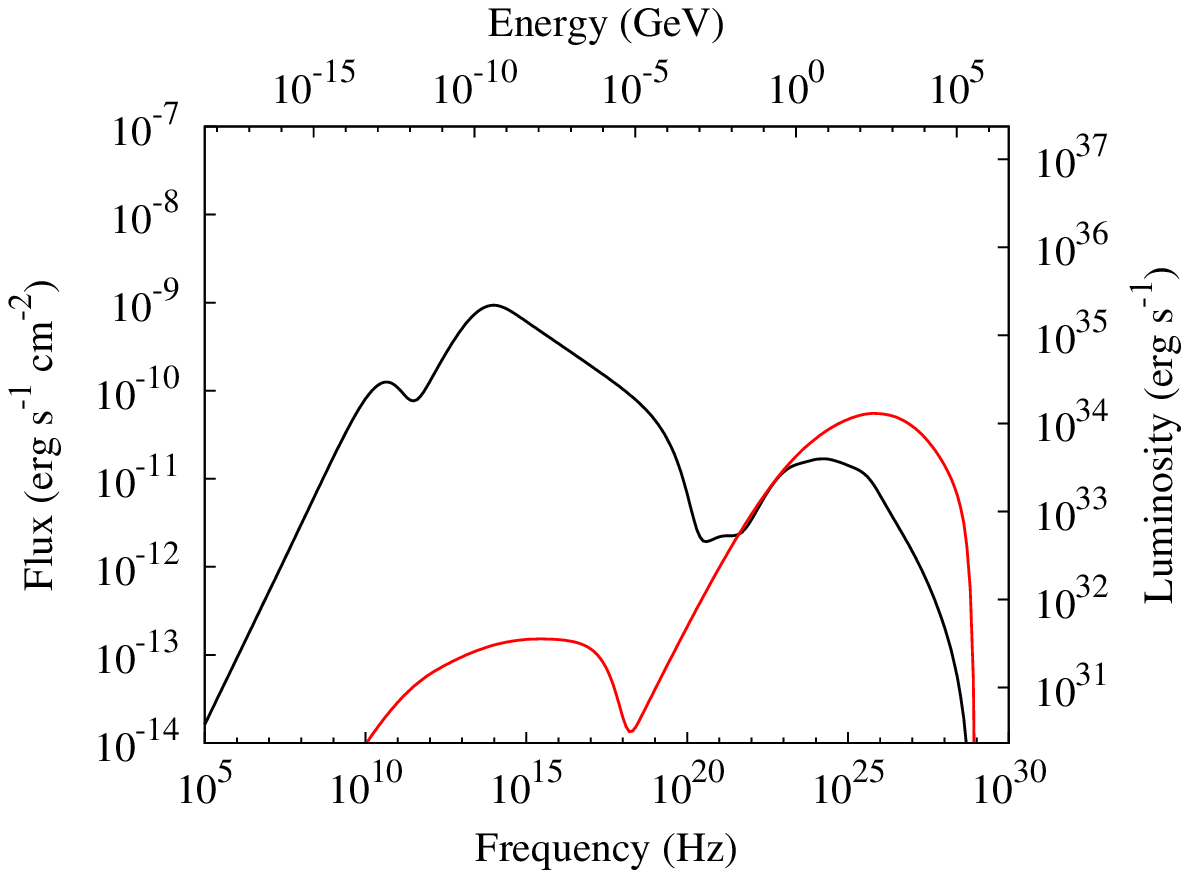}
\end{center}
\caption{Comparison between the pairs (left) and the PWN spectrum (right) obtained using the parameters shown in Table
\ref{simulation}, at 6, 10, 18, and 40 kyr (from top to bottom). The black and red lines show the case where we consider reverberation and 
only free expansion,
respectively. 
}
\label{comp}
\end{figure*}

During the reverberation phase, when there are evident effects on the dynamics (compression and bouncing of the nebula radius), there are also
important differences in the radiated PWN spectrum when compared to a freely expanding case. 
To briefly explore this for a PWN having a typical (and very low) instantaneous sharing $\eta$, in Table
\ref{simulation} we set the input parameters for a 40 kyr PWN. 
The evolution of the radius and the magnetic field for the
reverberation and the free expansion models of this nebula is shown in Fig. \ref{rad_comp}. 
We can observe that both quantities evolve in the same way
during the first $\sim$4 kyr (see in Fig. \ref{rad_comp}). Also the spectrum is similar.
The age at which the transition between different stages of the evolution occurs varies with the energy of
the SN explosion, the ejected mass, and the initial profiles of the SNR ejecta. 
At an age of $\sim$4 kyr, the PWN shell goes into the shocked medium of the remnant and starts the compression. 
During this phase, the
magnetic field and the internal pressure increases as they evolve approximately as $B(t) \sim R^{-3/2}$, $P \sim R^{-3}$. 
In order to see this first deviation, in Fig.
\ref{comp} we show the comparison of the pairs and the PWN spectrum at the age of 6 kyr. 
The increase of the magnetic field enhance
the synchrotron flux and cools down the high energy pairs ($\gamma \sim 10^8-10^{10}$) while  increasing slightly 
the population at $\gamma \sim 10^5-10^8$. 
The pairs at this range of energy interact with the IC target photon fields (CMB, FIR and NIR), thus also increasing
slightly the IC radiation flux between 1 GeV and 10 TeV. 
These effects become more evident as the compression goes on, as we can see at
10 kyr (second row of panels in Fig. \ref{comp}).

At the moment of maximum compression, the magnetic field is high enough to burn off the majority of high energy pairs and move them to
lower energies in the pair population spectrum. 
When the pressure is high enough too, the PWN bounces and starts re-expanding ($\sim$16 kyr). 
Then, the magnetic field in the nebula decreases again and the new high energy particles that are injected by the pulsar start accumulating. 
In Fig. \ref{comp},
we observe these effects at 18 kyr, where we see how the old high energy pairs are accumulated at the lower energy part of the pair
spectrum and the quantity of high energy particles is orders of magnitude lower in comparison to the free expansion model. 
The
PWN spectrum shows that the IC flux is two orders of magnitude lower after compression. 
The synchrotron flux is still higher
for the reverberation case, where the radius of the PWN is smaller and the magnetic field is high enough to produce a synchrotron flux
two orders of magnitude higher than in the free expansion model. Differences between models considering reverberation and not, are large, as expected (see \cite{gelfand2009}).

After the bounce (see the 40 kyr case in Fig. \ref{comp}), the radius of the PWN keeps growing freely until reaching the pressure of the Sedov
solution. At this stage, the PWN expansion is in equilibrium with the ambient medium and follows the Taylor-Sedov law.
The pulsar injects newly accelerated pairs that contributes to reestablish the high energy pair population. 
However,
we still note differences due to the extinction of high energy particles during the compression phase and the synchrotron losses, which
are higher than in the free expansion model, where the magnetic field is very low due to the unrealistic size of the PWN at this age if a free-expansion model is considered.
Note that at the same age after the bounce, considering the same instantaneous sharing, the synchrotron flux in the free expansion model may be
up to 3 to 4 (depending on the frequency) orders of magnitude lower than what we get if we include reverberation for the case we studied.

\subsection{An evolving energy partition}

\begin{figure}
\begin{center}
\includegraphics[width=0.45\textwidth]{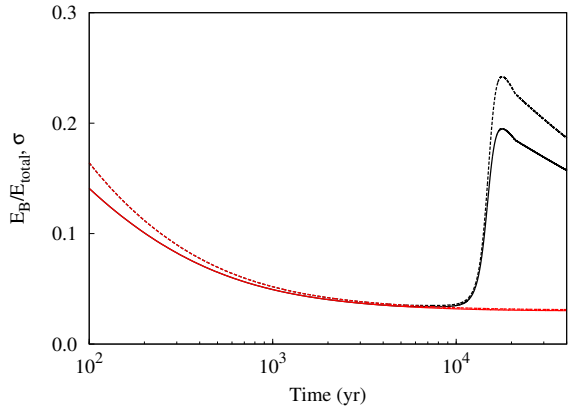}
\end{center}
\caption{Solid Lines: Ratio between the magnetic energy content to the total energy content (particle+field) for the generic model of the PWN
discussed in Section 2 ($E_B/E_{total}$), with parameters given Table 1. Dashed line: We also include the nebula-averaged 
$\sigma$-parameter defined as the ratio
between the magnetic field energy and the particles energy content only. In red, for a model considering only a free expansion; in black, for a model with reverberation
included.}
\label{mag-frac}
\end{figure}

Since the equation for the pressure evolution of the particle content in the nebula considers the particles's adiabatic and radiative losses, 
while the magnetic energy follows the adiabatic law, 
the ratio of particle's energy to magnetic field energy, is no longer constant in time, neither it is equal to   $1-\eta$. 
However, the modelling provides $B(t)$, as well
as the particles's energy via Eq. (\ref{epar}), and thus the energy ratio can be computed directly. This is shown in Fig. \ref{mag-frac},
where we plot the ratio of $E_B$ with the total energy $E_B + E_p = E_{\rm total}$ and with the particle's energy (the averaged $\sigma$ parameter).

Due to the high-energy losses at the beginning of the evolution, the ratio between the magnetic energy and the total energy content of the nebula is high with respect the parameter $\eta$ (the free parameter for the instantaneous sharing, as explained in the text above) used during injection. During the free expansion evolution of the nebula, the energy partition approaches to $\eta$, but it is still larger than it (what can be understood due to the fact that particles loose energy via radiation, and the field does not).
In compression, then, the increase of the magnetic field makes the energy partition to significantly deviate from the free expansion case.

\section{Modelling of the PWN in CTA 1}

\begin{table}
\vspace{4cm}
\centering
\scriptsize
\vspace{0.2cm}
\caption{Assumed or observationally constrained model parameters for the PWN in  CTA~1.}
\label{cta1_par1}
\begin{tabular}{lll}
\hline
Parameter & Value & Reference\\
\hline
\hline
Pulsar magnitudes\\
\hline
$P$ (s) & 0.316 & \citet{abdo2012}\\
$\dot{P}$ (s s$^{-1}$) & 3.6 $\times 10^{-13}$ & \citet{abdo2012}\\
$\tau_c$ (kyr) & 13.9 & \citet{abdo2012}\\
$L(t_{age})$ (erg s$^{-1}$) & 4.5 $\times 10^{35}$ & \citet{abdo2012}\\
$n$ & 3 & Assumed\\
\hline
PWN magnitudes\\
\hline
$d$ (kpc) & 1.4 & \citet{pineault1993}\\
$\gamma_{pwn}$ & 4/3 & Assumed\\
\hline
SNR magnitudes\\
\hline
$\gamma_{snr}$ & 5/3 & Assumed\\
$\omega$ & 9 & Assumed\\
$E_{sn}$ (erg) & $10^{51}$ & Assumed\\
\hline
Photon environment\\
\hline
$T_{cmb}$ (K) & 2.7 & Assumed\\
$w_{cmb}$ (eV cm$^{-3}$) & 0.26 & Assumed\\
$T_{fir}$ (K) & 25 & GALPROP\\
$T_{nir}$ (K) & 3000 & GALPROP\\
\hline
Injection parameters\\
\hline
$\gamma_{min}$ & 1 & Assumed\\
$\gamma_b$ & $10^{6}$ & Assumed\\
$\alpha_l$ & 1.5 & Assumed\\
$\epsilon$ & 0.3 & Assumed\\
\hline
\hline
\end{tabular}
\end{table}

\begin{figure*}
\begin{center}
\includegraphics[width=0.35\textwidth]{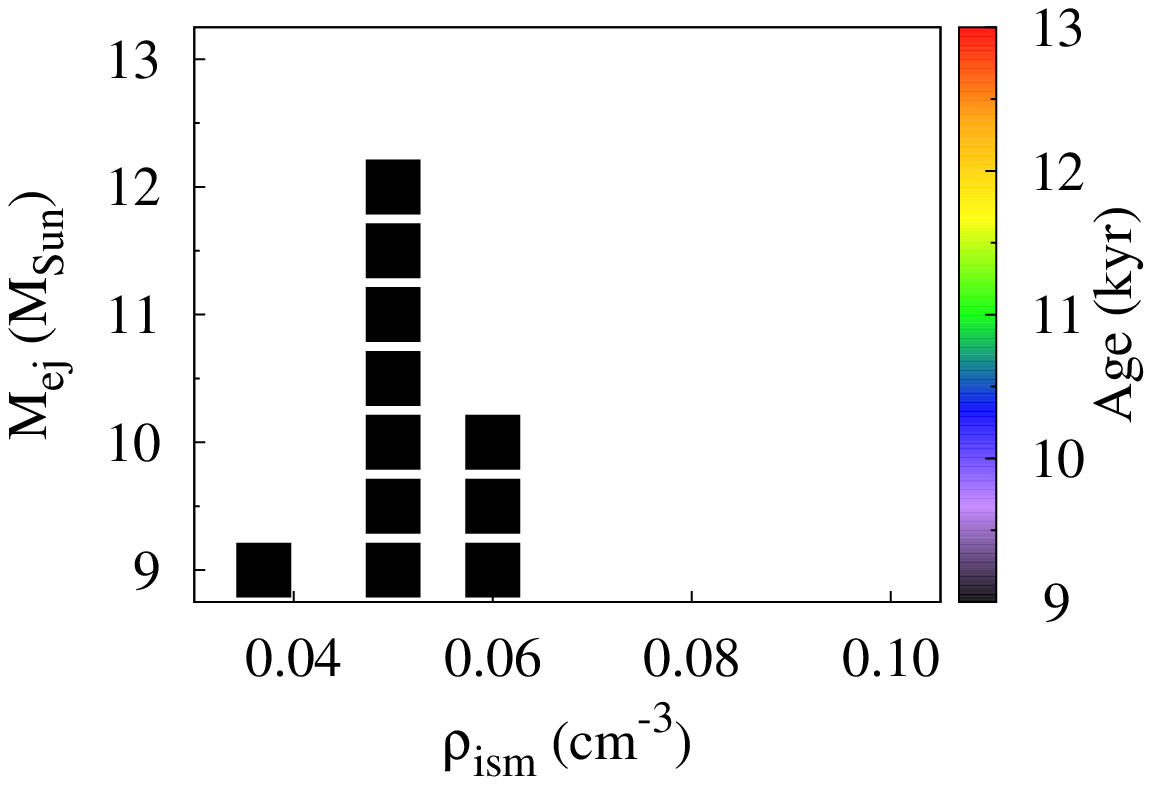}
\includegraphics[width=0.35\textwidth]{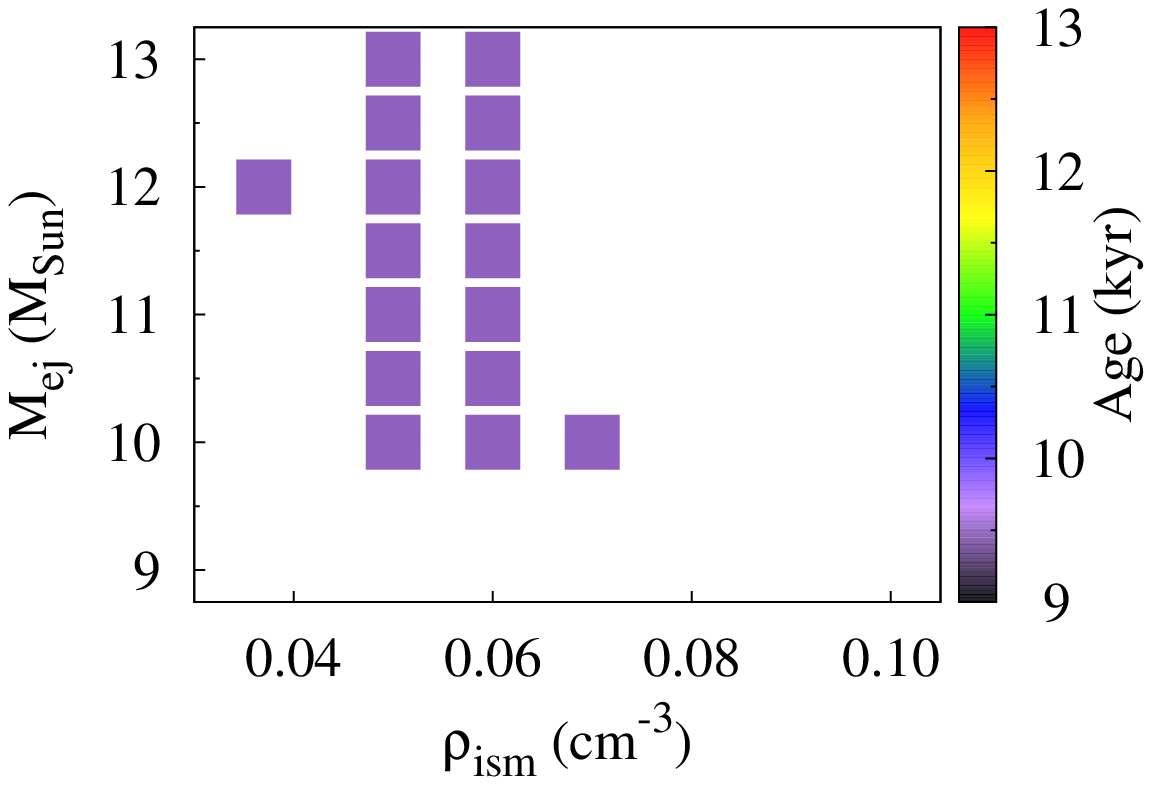}
\includegraphics[width=0.35\textwidth]{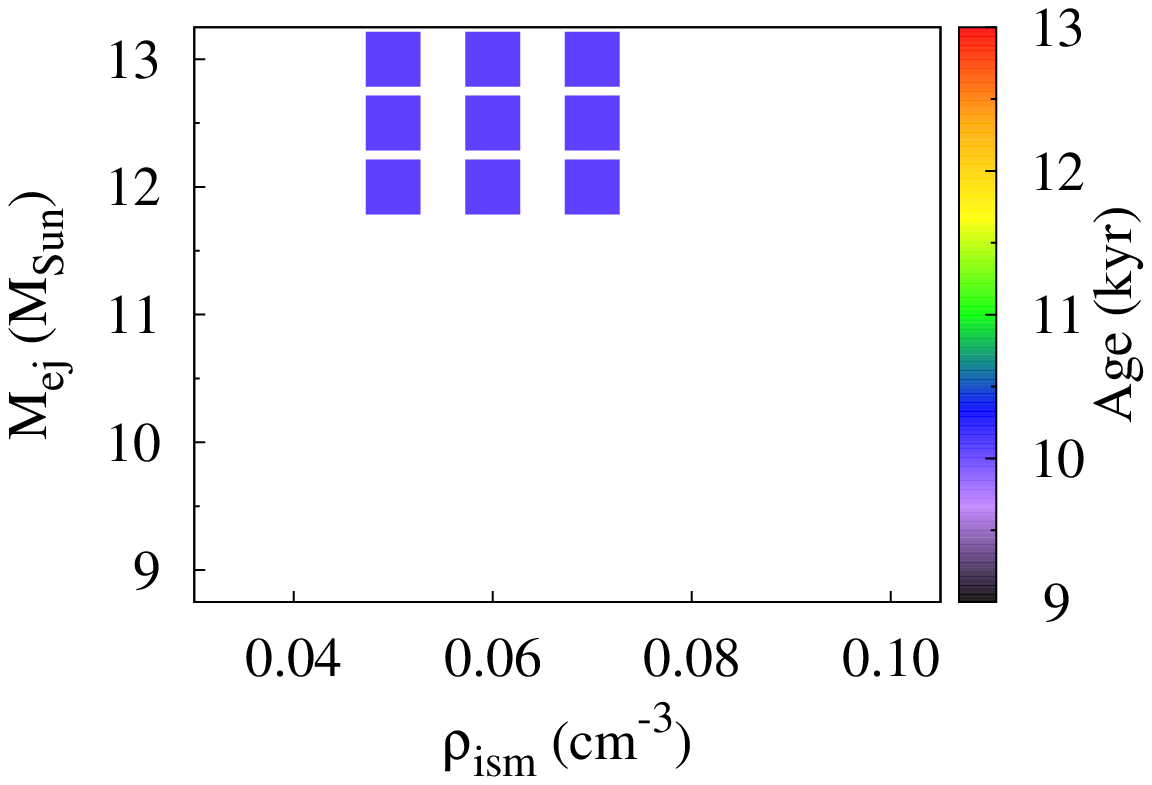}
\includegraphics[width=0.35\textwidth]{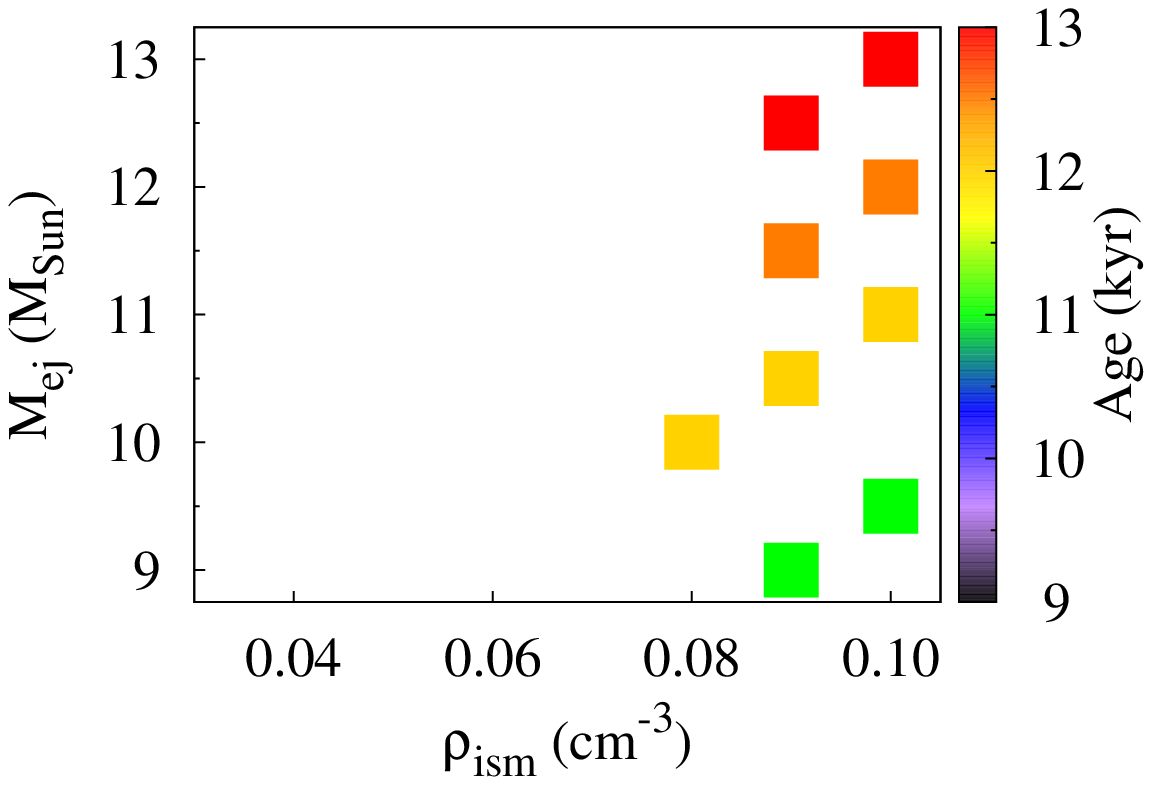}
\end{center}
\caption{Models that result compatible with the observational values of the radii of the CTA 1 SNR and PWN. The two plots on the top and the left plot on
the bottom row show models in free expansion. We show them separated in three different plots to avoid superposition of the points. The
appearance of models with the same ejected mass and ISM density but with different ages is due to the uncertainty on the CTA 1 distance and the
tolerance that we allow to fit the radius of the PWN. The right plot on the bottom correspond to models in compression phase.}
\label{colourmaps}
\end{figure*}

We shall first consider the constraints that the SNR and PWN radius impose on models.

The CTA~1 radio SNR has a radius of $\sim 20.4d_{1.4}$ pc \citep{pineault1993}; therefore, taking into account the uncertainty on the
distance, we will accept models able to cope with a constraint in the SNR radius of  $19.5 < R_{snr} < 21.5$ pc. 
On the other hand, our model (being 1D) assumes
a spherical system. 
We take as a reference the dimensions of CTA~1 PWN described by {VERITAS}
($7.3d_{1.4} \times 5.9d_{1.4}$ pc, \citealt{aliu2013}), and we will consider acceptable 
any value for the PWN radius within this range.

The radius of the SNR and the PWN are determined by three parameters: the ejected mass $M_{ej}$, the age the system $t_{age}$,
and the interstellar medium (ISM) density $\rho_{ism}$. 
The rest of the assumed parameters (including those taken directly from observations) related with
the evolution are quoted in Table \ref{cta1_par1}. 
We consistently computed the initial spin-down luminosity and the spin-down age.
We explore models that fit the radii assuming a range of  ejected masses between 9 and 13$M_\odot$ (in the usual range for a SN explosion with a typically massive progenitor, note instead that
Aliu et al. (2013) have assumed a lower ejected mass of 6.5 $M_\odot$) and
ages that range from 9 kyr  to 13 kyr \citep{slane2004}. We have computed $\sim$650 spectra varying the assumed {\it average} density 
from 0.05 cm$^{-3}$ to 0.1 cm$^{-3}$. The latter value is the
upper limit for the ISM density of the Galaxy at the CTA 1 position \citep{dickey1990,slane1997}.
We have considered too the observational indications for a density of 0.017  \citet{slane2004} 
and 0.037 cm$^{-3}$ \citet{slane1997} in particular regions. In any case, all these values are still much lower than the upper limits we have
found above from our analysis of molecular material in the vicinity.

%
Depending on the values of $M_{ej}$,
$t_{age}$, and $\rho_{ism}$, we have different compatible models fitting the constraints for the radii, given the observational uncertainties.
Their spectral features are different, since the PWN would be in a different stage of evolution (free expansion or compression), and
we discuss them next. Our first result is that no models in the late Sedov expansion phase, within the explored parameter ranges mentioned above, fit the observed radii.

In Fig. \ref{colourmaps}, we show those models that do fit the observational values for the radii of the SNR and the PWN. 
Note that some models have, e.g., similar
ejected mass and ISM density, but the age of the system is different. In order to see the variety of these models, Fig. \ref{colourmaps}
shows three plots for those models which are in free expansion (see the caption for a detailed explanation).
%
The models in free expansion fit the spectrum with instantaneous sharings between 0.25 and 0.4. An upper limit for the FIR and NIR densities
is established at 0.037 and 0.075 eV cm$^{-3}$, respectively. These values are lower than those obtained from
GALPROP\footnote{\url{http://galprop.stanford.edu/}} \citep{porter2006}, which predicts $w_{fir}=0.3$ eV cm$^{-3}$ and $w_{nir}=0.6$ eV
cm$^{-3}$ for the Galactic position of CTA 1, and can be understood in the context of local fluctuations.

For models in compression
and those cases in which the constraints in the radii are fulfilled, as shown in Fig. \ref{colourmaps},
we do not get an acceptable fit for the TeV flux. In all these cases,  the IC flux due to the CMB target field would already exceed the observational constraints. 
This can be understood by the accumulation of cooled electrons with Lorentz factors between $\gamma \sim 10^5$ and $ 10^8$, as
shown in Fig.~\ref{comp}.

\begin{figure*}
\begin{center}
\includegraphics[width=0.45\textwidth]{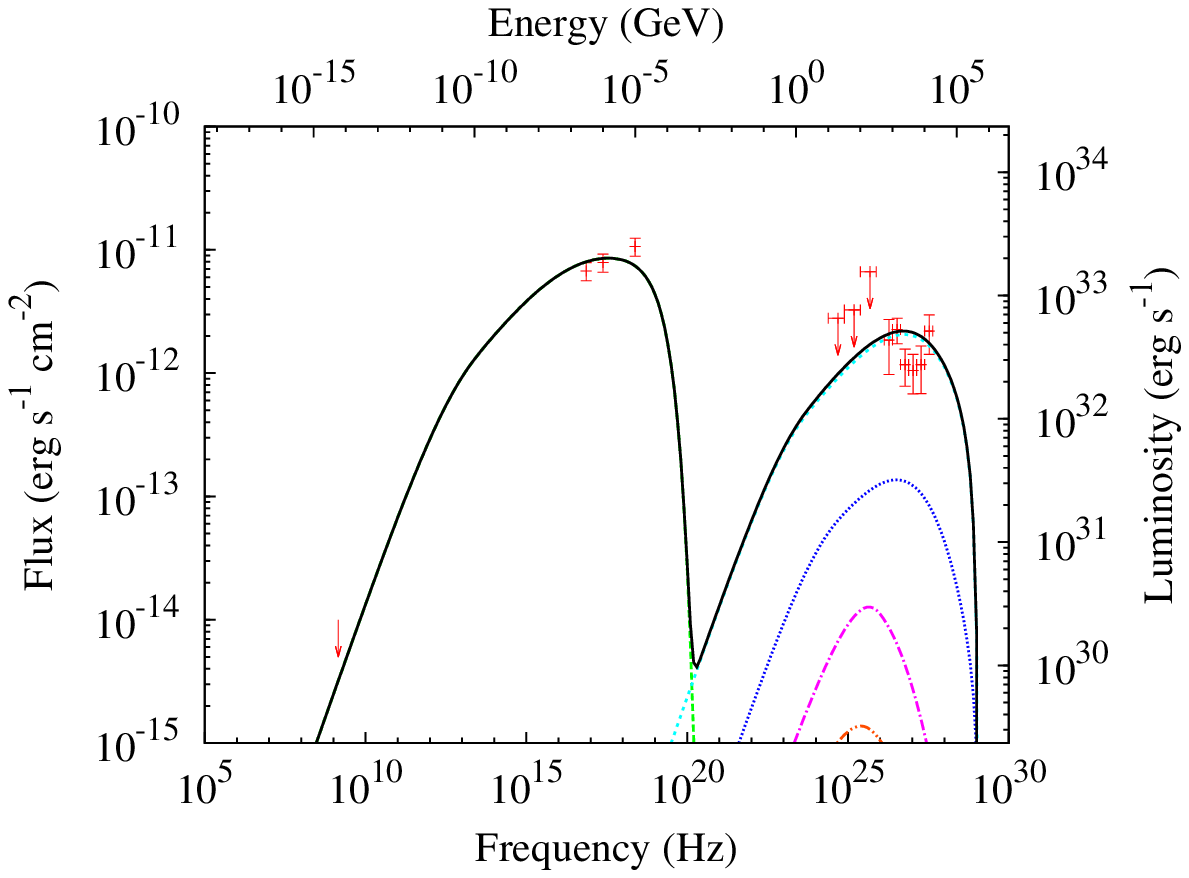}
\includegraphics[width=0.45\textwidth]{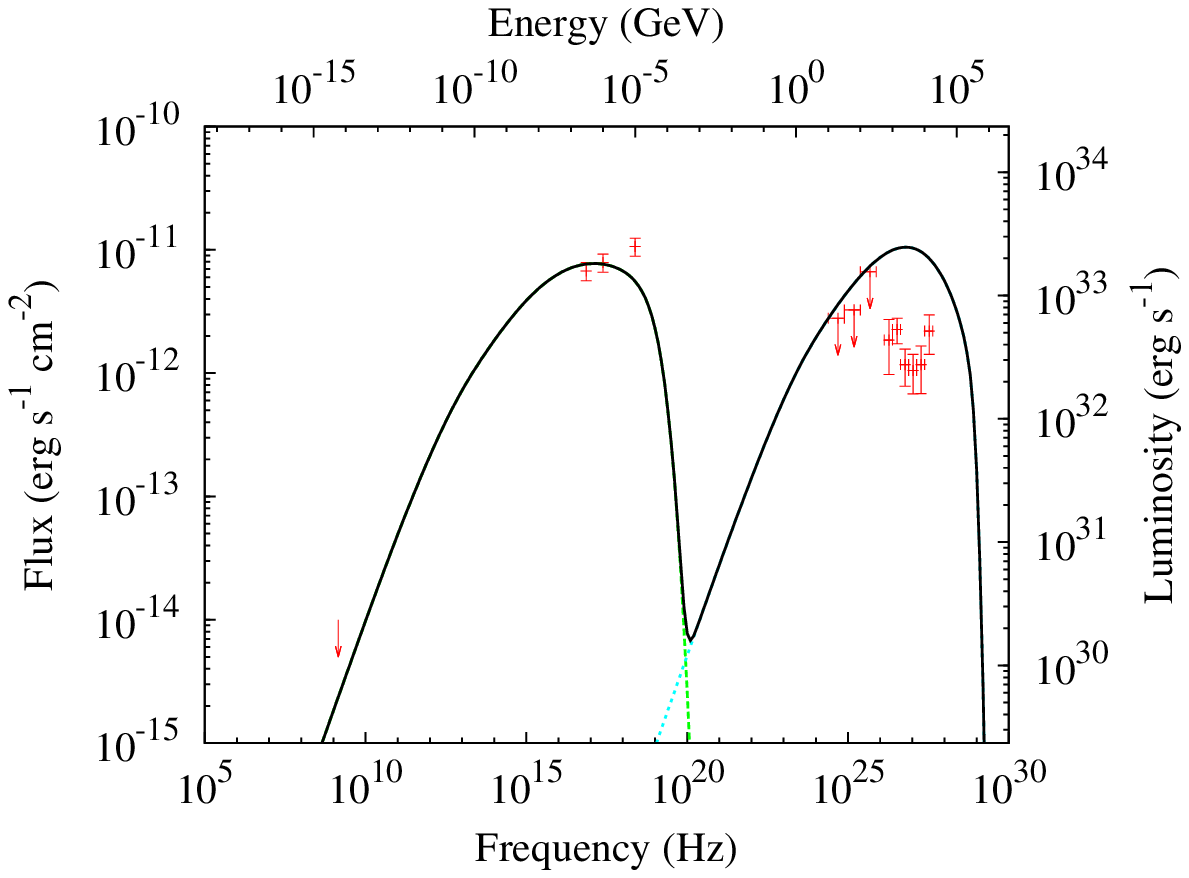}\\
\end{center}
\caption{Two models for CTA 1. In free expansion (left), a high instantaneous sharing parameter of  $\sim$0.25 is needed to achieve the X-ray fluxes and the
energy densities for the FIR and NIR target fields need to be low (see Table \ref{cta1_par2}). In compression (right), only the CMB IC flux
contribution exceeds the {VERITAS} data points at VHE. The color legend is the following: green long-dashed line corresponds the synchrotron
emission, cyan dashed line to IC-CMB emission, blue dotted lines to IC-FIR emission, purple dot-dashed line to IC-NIR emission and orange doubble
dotted-dashed line to Bremsstrahlung emission. SSC component is out of the scale of the plot and negligible. Data points in radio come from
\citet{giacani2014}, in X-rays from \citet{lin2012}, in VHE from \citet{aliu2013}. The {\it Fermi} upper limits are given in \citet{acero2013}.}
\label{cta1_models}
\end{figure*}
\begin{table*}
\vspace{4cm}
\centering
\scriptsize
\vspace{0.2cm}
\caption{Parameters obtained from the fitted spectrum for CTA 1 PWN. Note that the model in compression phase does not fit the observational data, but it is shown as
an example for the discussion.}
\label{cta1_par2}
\begin{tabular}{lll}
\hline
Parameter & Value\\
\hline
\hline
Pulsar Wind Nebula Magnitudes & Free expansion & Compression\\
\hline
$t_{age}$ (kyr) & 9.2 & 11.4\\
$L_0$ (erg s$^{-1}$) & $3.9 \times 10^{36}$ & $1.4 \times 10^{37}$\\
$\tau_0$ (kyr) & 4.7 & 2.5\\
$v_{pwn}$ (km s$^{-1}$) & 808 & -6680\\
$R_{pwn}$ (pc) & 6.7 & 6.7\\
$M_{sw}$ (M$_{\odot}$) & 0.1 & 0.11\\
$P_{pwn}$ (Ba) & $1.4 \times 10^{-12}$ & $2.1 \times 10^{-12}$\\
$E_{pwn}$ (erg) & $10^{47}$ & $2.2 \times 10^{47}$\\
\hline
Supernova Remnant magnitudes\\
\hline
$M_{ej}$ (M$_{\odot}$) & 12 & 9\\
$\rho_{ism}$ & 0.037 & 0.08\\
$R_{snr}$ (pc) & 21.1 & 20.8\\
$R_{rs}$ (pc) & 15.2 & 8.6\\
\hline
Photon environment\\
\hline
$w_{fir}$ (eV cm$^{-3}$) & 0.037 & 0\\
$w_{nir}$ (eV cm$^{-3}$) & 0.075 & 0\\
\hline
Injection parameters\\
\hline
$\gamma_{max}(t_{age})$ & $10^{9}$ & $8.0 \times 10^{8}$\\
$\alpha_h$ & 2.3 & 2.1\\
\hline
Magnetic field \\
\hline
$B(t_{age}) ({\mu}G)$ & 4.3 & 1.8\\
$\eta$ & 0.25 & 0.017\\
\hline
\hline
\end{tabular}
\end{table*}

Fig. \ref{cta1_models} shows a free expansion model for CTA 1 that fits the spectrum, and a compression stage model that does not, as an example. The
parameters for these models are summarized in Table \ref{cta1_par2}. The fitted parameters are mainly those related with the particle injection,
photon environment, and  instantaneous sharing fraction. Some other parameters in the table are derived, using model equations (see Table \ref{cta1_par1}).
Note that the free expansion model considers an ISM density of 0.037 cm$^{-3}$. 
At $\rho_{ISM}=0.017$ cm$^{-3}$, all models have a SNR radius larger than 21.5 pc. The free expansion model
fitting that we show comes from the refinement of the parameters in order to fit an ``average'' radius for CTA 1 PWN of 6.7 pc. Taking this into
account, we get only one model as viable, from the two shown in Fig. \ref{colourmaps} for $\rho_{ISM}=0.037$ cm$^{-3}$, with an ejected mass of 12$M_\odot$
and an age of 9.2 kyr.

\begin{figure*}
\begin{center}
\includegraphics[width=0.45\textwidth]{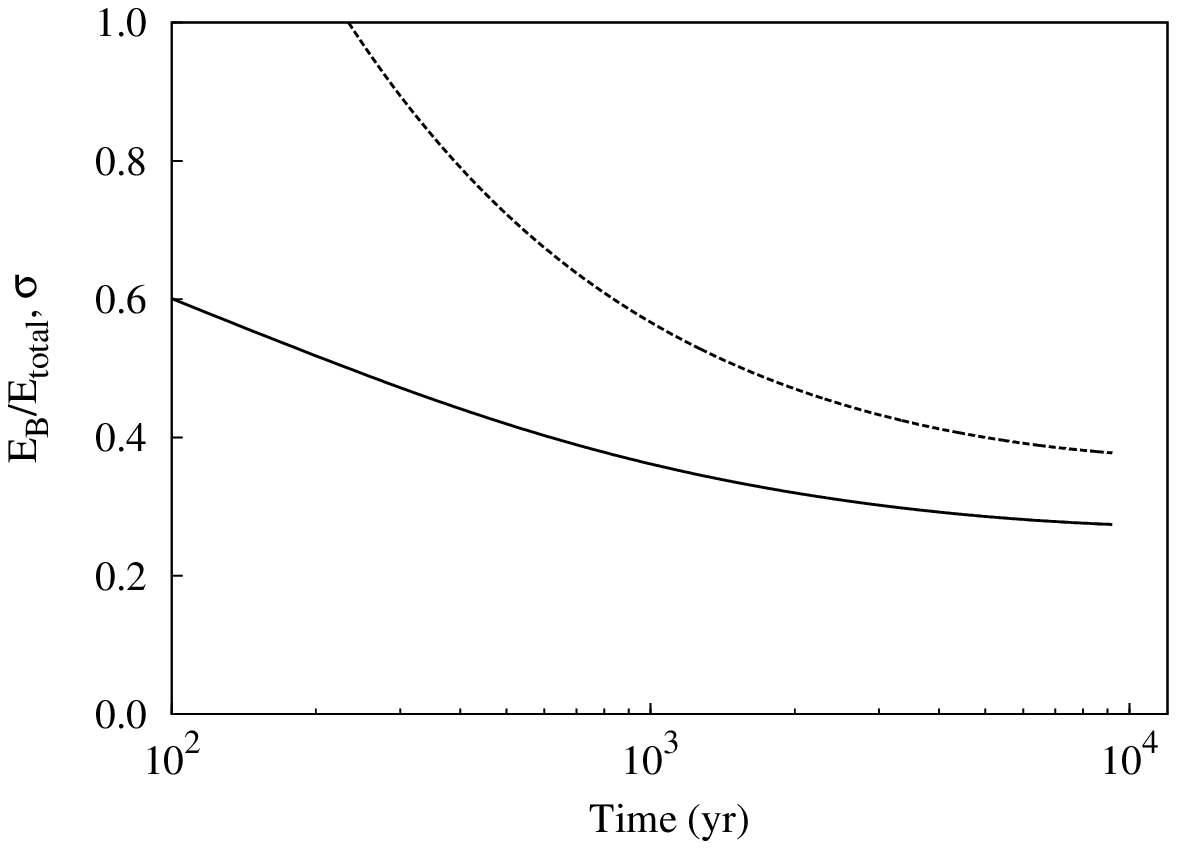}
\includegraphics[width=0.45\textwidth]{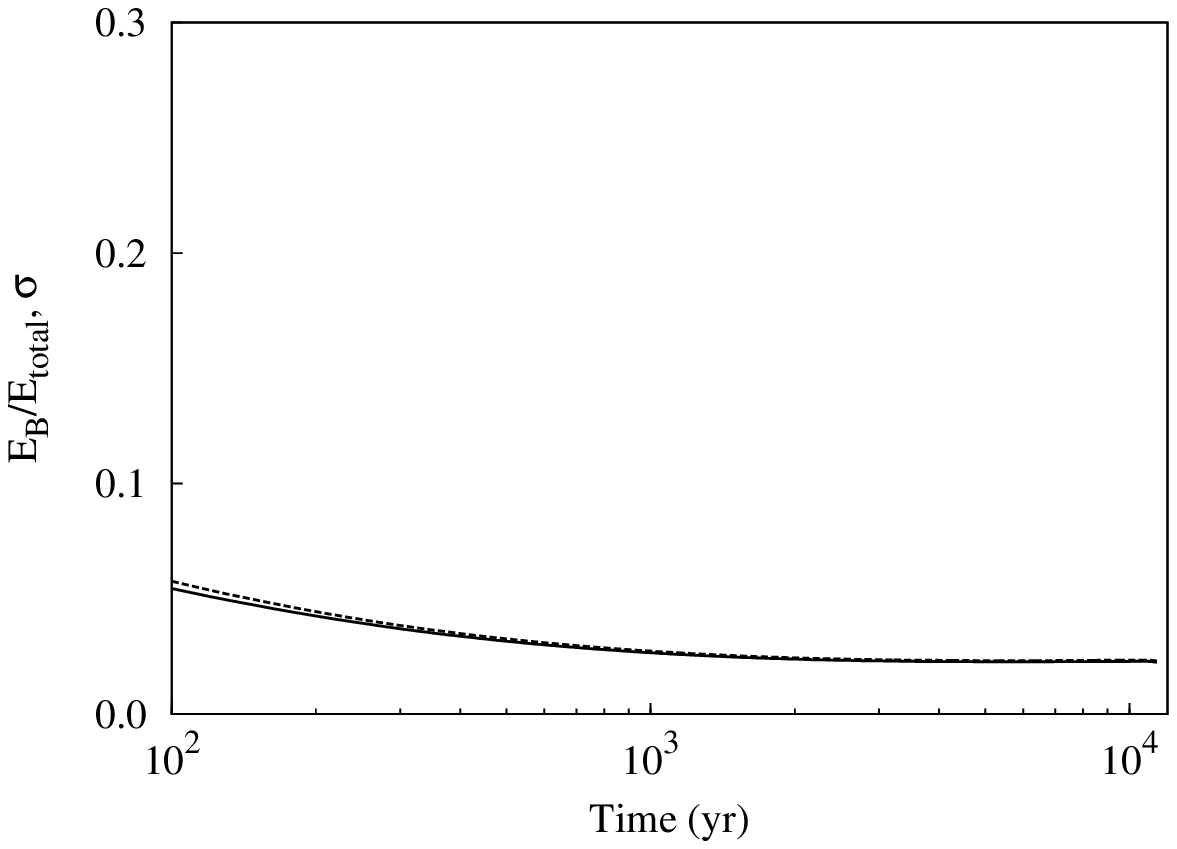}
\end{center}
\caption{Ratio between the magnetic energy content to the total energy content (particle + field) for the CTA 1 PWN model. We
include the $\sigma$-parameter in a dashed line (hardly visible in the right plot). Left panel: free expansion model with an age of 9.2 kyr; right panel: (compression model at an age of 11.4 kyr; the change in these ratios due to compression (as shown above in Fig. 4 is still not visible, but would show up similarly, and well before 20 kyr, if we let this model continue run in time.
For more details see text.}
\label{mag-cta}
\end{figure*}

Regarding the evolution of the energy partition in the CTA nebula, we show it in Fig. \ref{mag-cta} for the two models considered above, as an example.

\subsection{Discussion}

We see that the observational data for the CTA~1 PWN is problematic, and cannot be satisfactorily fitted by our
relatively-detailed model without considering 
important caveats. 

If we consider a fit during the free expansion phase, the instantaneous sharing (and the 
energy partition of the nebula) needs to be relatively high (in comparison to all other young PWNe, see \citealt{torres2014}). This is needed 
in order to obtain the appropriate magnetic field to fit the X-ray data with synchrotron radiation. If the PWN is in a compression state, the problem arise from
over-predicting the high energy fluxes, as shown above.

\begin{figure*}
\begin{center}
\includegraphics[width=0.45\textwidth]{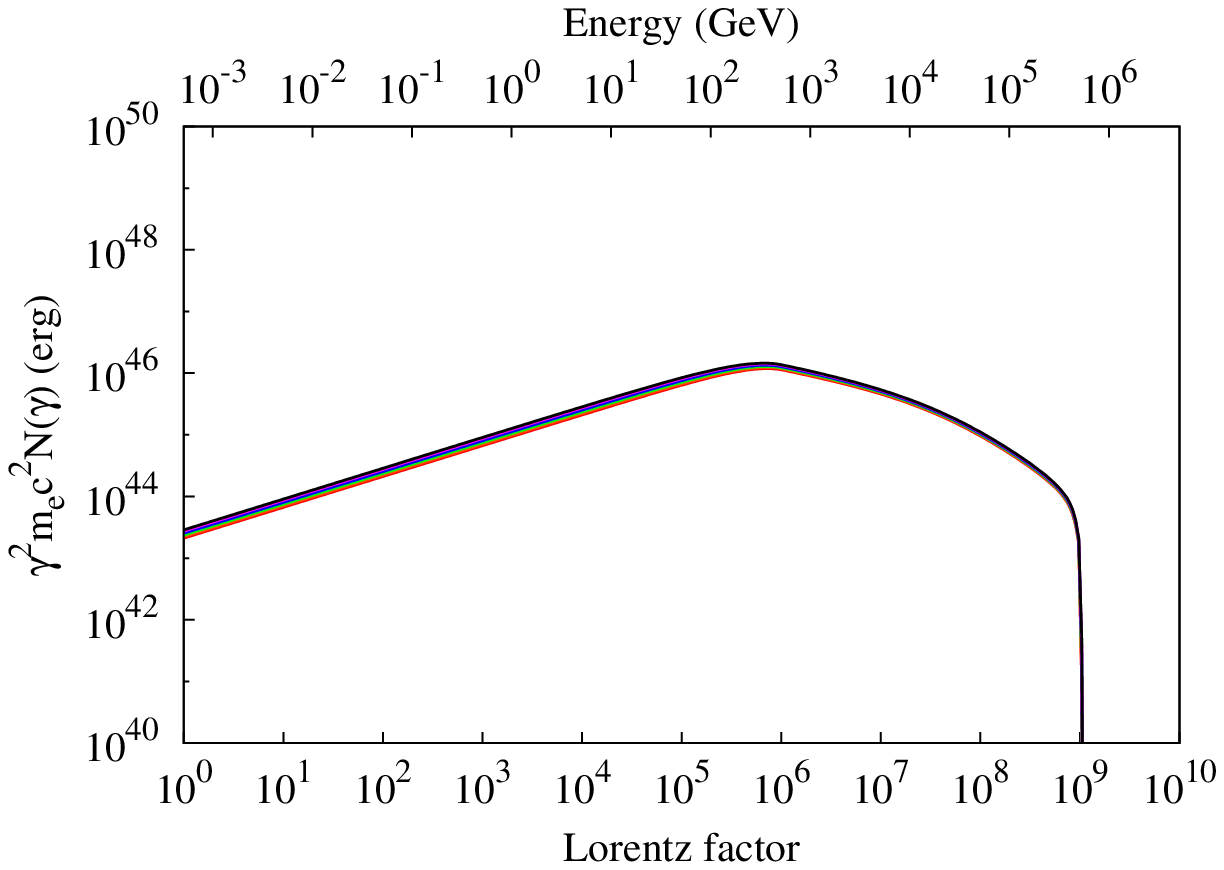}
\includegraphics[width=0.45\textwidth]{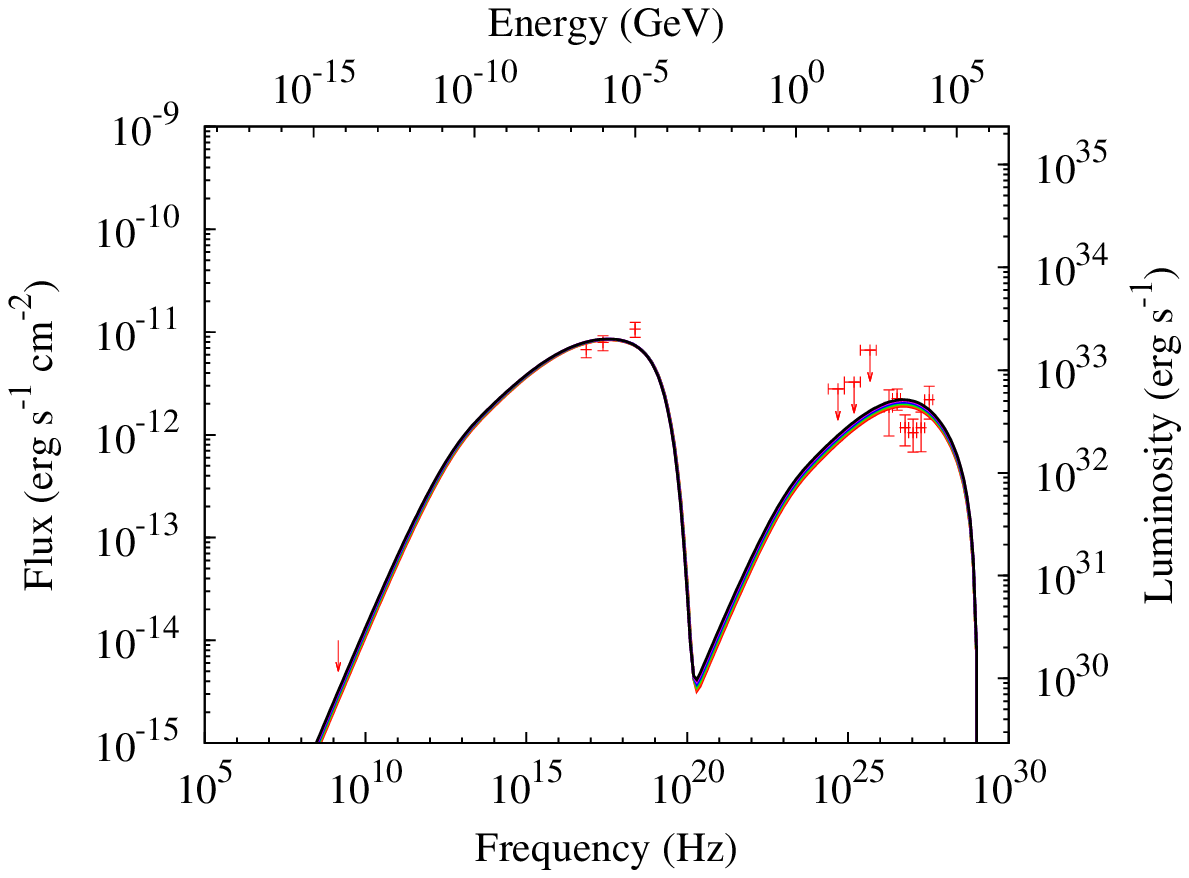}\\
\includegraphics[width=0.45\textwidth]{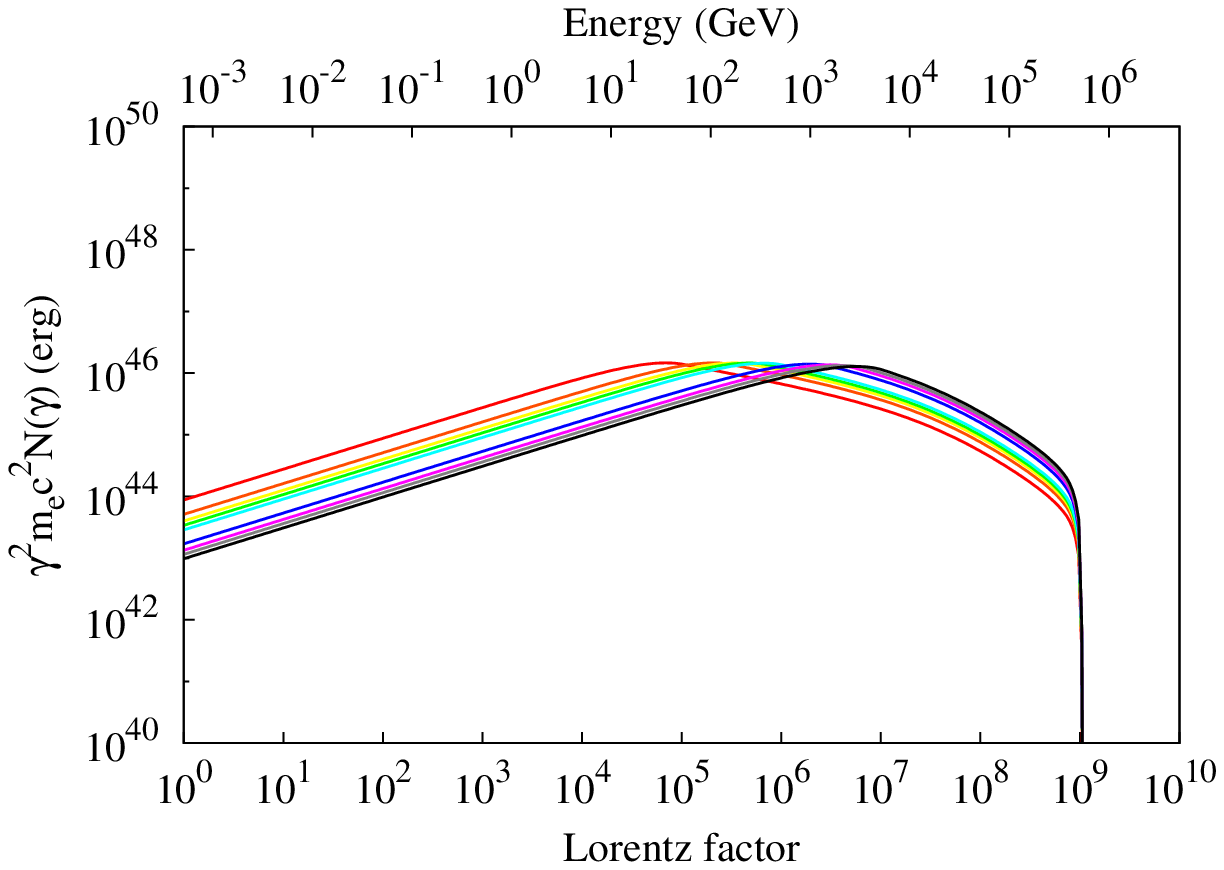}
\includegraphics[width=0.45\textwidth]{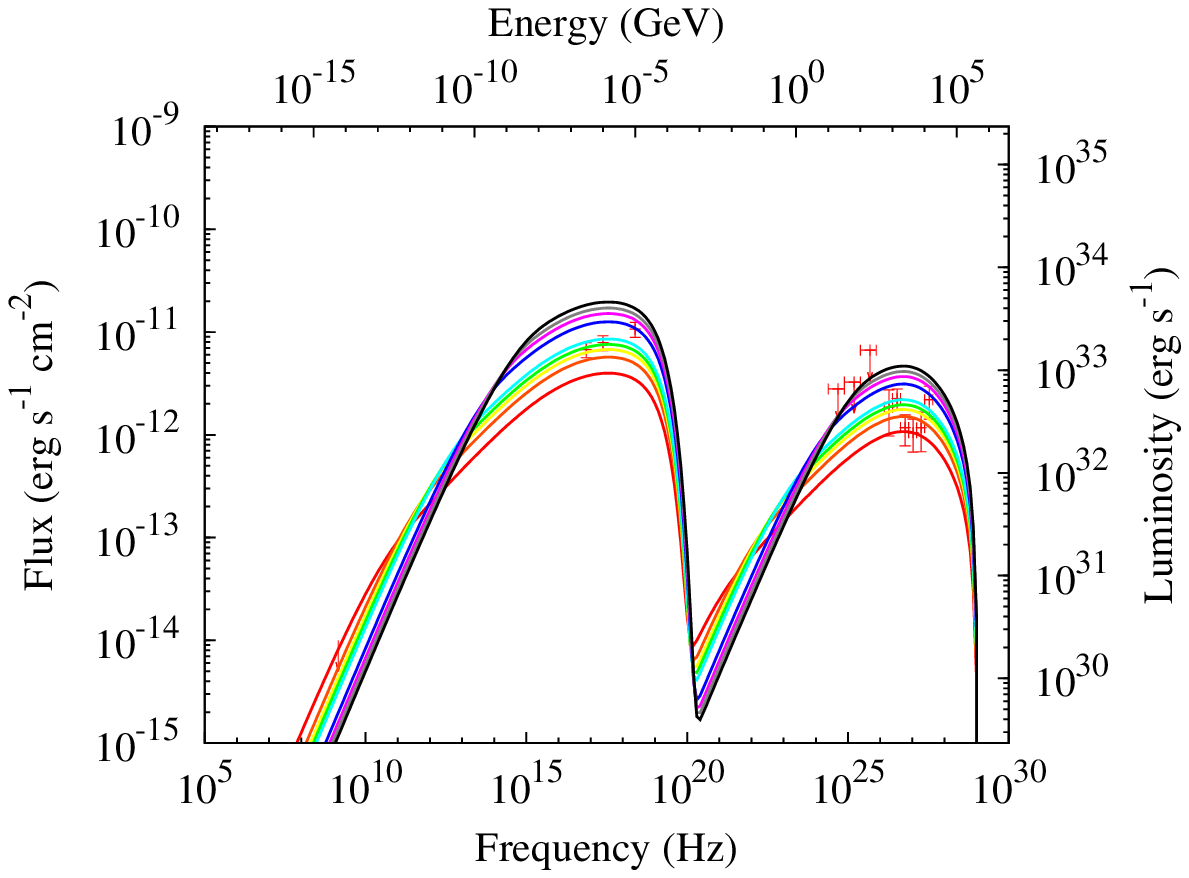}\\
\includegraphics[width=0.45\textwidth]{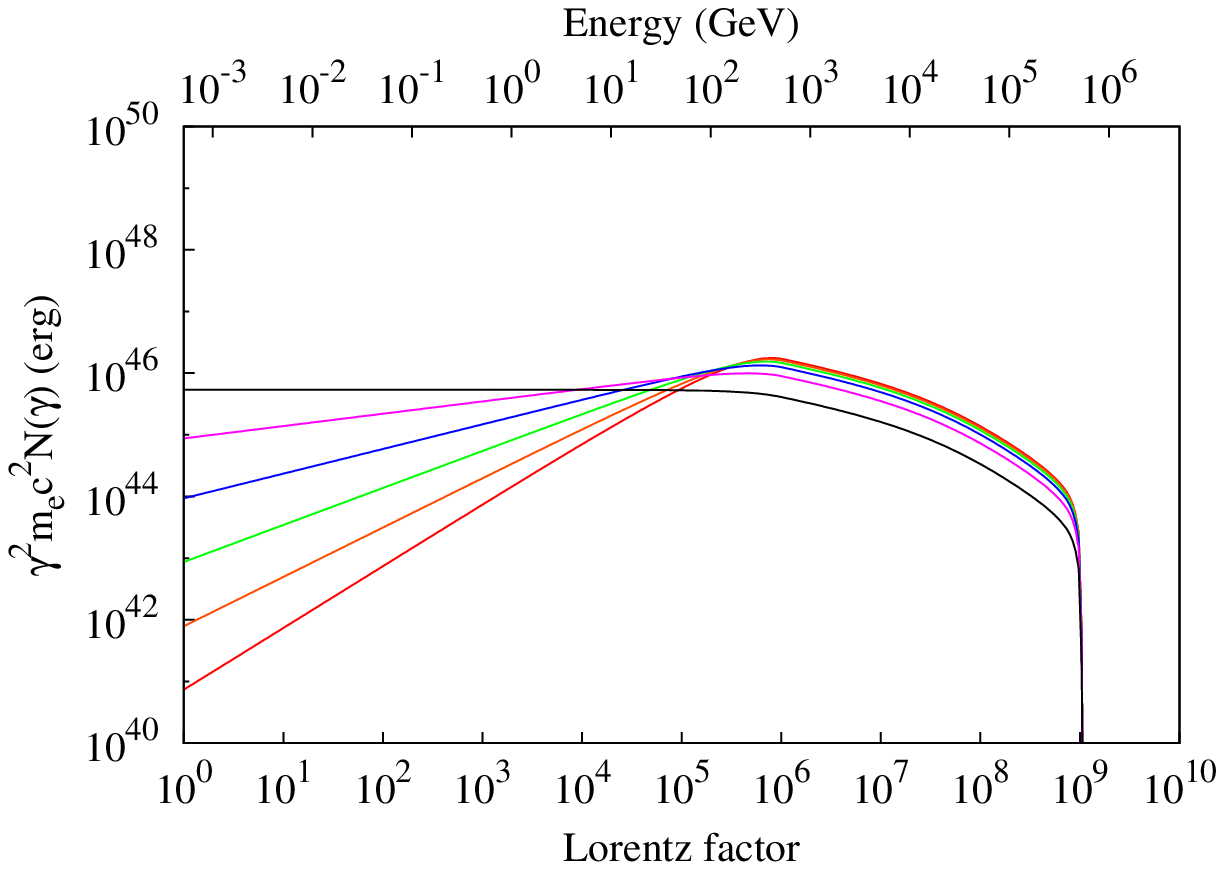}
\includegraphics[width=0.45\textwidth]{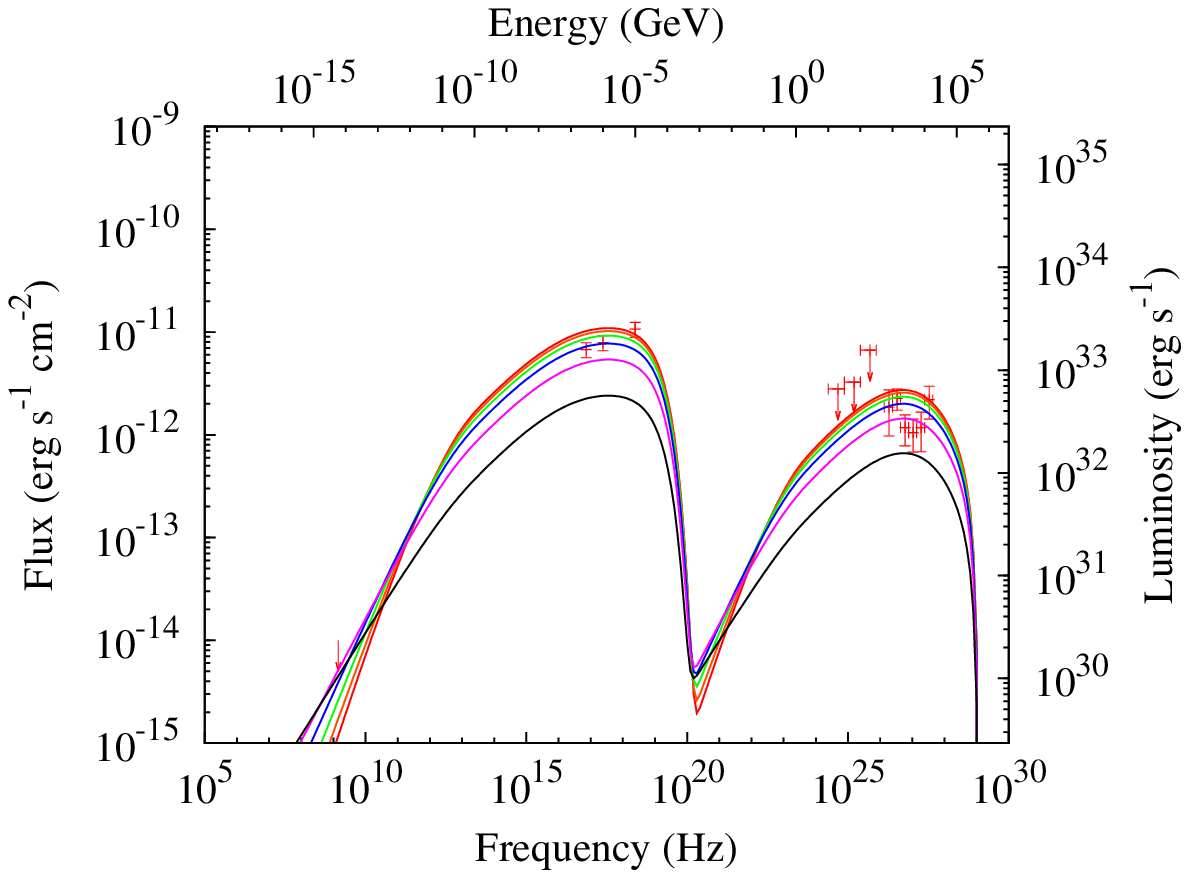}
\end{center}
\caption{Variation of the electron (left) and spectral (right) model for CTA 1 with the braking index $n$ (upper panels), energy break $\gamma_b$
(middle panels) and low energy spectral index $\alpha_l$ (lower panels). The color legend is: for upper panels, the red, orange, green, blue,
purple and black lines correspond to $n=2.1,2.3,2.5,2.7,2.9,3$, respectively; for middle panels, the red, orange, yellow, green, cyan, blue,
purple, grey and black lines correspond to
$\gamma_b=10^5,3 \times 10^5,5 \times 10^5,7 \times 10^5,10^6,3 \times 10^6,5 \times 10^6,7 \times 10^6,10^7$; and for lower
panels, the red, orange, green, blue, purple and black lines correspond to $\alpha_k=1,1.2,1.4,1.6,1.8,2$.}
\label{cta1_var}
\end{figure*}

We recall that additional considerations done in our simulations are the fixing of the braking index as $n=3$, the energy break as
$\gamma_b=10^6$, and the low energy index as $\alpha_l=1.5$. The latter two parameters are  unconstrained, 
since at
low energies in the spectrum, we only have an upper limit on the radio flux from the estimation done by \citet{giacani2014}.
The values we have assumed for these parameters 
are typical \citep{bucciantini2011,torres2014}. However,
in Fig. \ref{cta1_var} we show the results of variations in these parameters taking as a reference model
the fit for the PWN in free expansion that we show in Table \ref{cta1_par2}. In the upper panels of Fig. \ref{cta1_var}, we show the variation
of the spectrum with the braking index going from 2.1 to 3. Note that the spectrum is almost independent on this parameter. In the middle
panels, we show the variation with the energy break from $\gamma_b=10^5$ to $10^7$. Note that models with $\gamma_b > 10^6$ exceeds the
observational synchrotron and IC flux. The synchrotron flux can be corrected by decreasing the instantaneous sharing
(which plays in favour to the
tendency that we observe in \citealt{torres2014}), but the IC flux generated considering only the CMB contribution is too high. 
Note also that those models with
$\gamma_b < 10^6$ could fit better the IC flux using the FIR and NIR energy densities given by GALPROP, but the synchrotron X-ray flux is too
low and we would need to increase the instantaneous sharing even more, up to the point to exceed the upper limit in the radio band. Regarding the lower
panel, we see a similar behaviour when we change the value of the low energy index $\alpha_l$. We vary this parameter from 1 to 2. 
We have also varied
the slope of the density describing the ejecta, from $\omega$=7 to 13, finding that the differences with our reported results are relatively minor.

From a modelling perspective, the age of the PWN as well as its morphology does not favor it being in a pure free expansion phase,
despite the important caveats that we have seen for
the compression phase in our models.
A way out could be that the cooling of pairs during compression  is different from what we have been able to model, 
due perhaps to a radial dependence of the magnetic field
(e.g., \citealt{gaensler2006}). 
If  so happens, particles in different regions of the PWN would not be affected by synchrotron energy losses in the same way, changing
the resulting pair population. 
An asymmetric evolution of the PWN due to e.g., different ISM densities, could explain this scenario. In this situation, one can imagine that 
some of the pairs would effectively evolve as in a freely
expanding nebula, and some others would be affected by the compression of the SNR reverse shock. 
An intermediate situation where the overall
IC yield is not increased beyond the measurements while an increased magnetic field reduce the need of a larger magnetization can be imagined. 
However,  this exploration is beyond the ability of our numerical scheme and would probably require a morphological/radiative approach to be consistently solved at once.
Finally, in terms of our model, the nebula cannot be re-expanding (Sedov phase) after compression.

\section{Concluding remarks}

We have found that the TeV emission from the direction of CTA 1 is very unlikely related to hadronic illumination of a molecular cloud in the
vicinity of the SNR. Planck and CO Harvard survey data are inconsistent with this interpretation. Thus, understanding the multifrequency
emission with a PWN model is important. 

We have included the prescription given by \citet{chevalier2005,gelfand2009} to solve the Euler equations of the PWN during the reverberation
phase into the radiative model described in \citet{torres2014}, and use it to study the effects of the compression and re-expansion of a
PWN due to the interaction with the shocked medium of the SNR. 
We have included the energy radiated by particles and their escape when computing the pressure produce by the gas. This leads to an evolving
energy partition, since for the same instantaneous sharing (magnetic fraction) of the injection of energy provided by the rotational power, the field and the particles are affected differently by radiation and losses.

We observed that there are significant effects in the spectrum compared with a free expansion model that induce differences in flux predictions
at all energies of one or two orders of magnitude. 
We note that with the same instantaneous sharing, models that take into account reverberation have more synchrotron flux during the bounce than
models considering  the nebula to be in free expansion at the same age. 
This may severely affect the determination of the instantaneous sharing $\eta$, and ultimately, of the energy partition.

We have applied this more complete radiative/dynamical model to CTA 1, which is a critical case due to its age (in the high end of those
considered to be young PWNe) and atypical because of its requirement of a relatively larger magnetization when free expansion models are
considered to fit the spectrum. 
We have made a parameter space exploration for the ejected mass, age, and ISM density that would give SNR and PWN radii as measured, and
consider those  in order to fit the CTA1 PWN spectrum. 
We could not find any completely acceptable representation at low values of magnetization, in any phase. 
Free expansion models (the only ones that can actually fit the data in the explored phase space) need instantaneous sharing parameters that are more than one order of magnitude larger
than the ones typically obtained for young PWNe. 
Compression models exceed the IC flux data even only considering the CMB target photon field.
And there are no models in a late Sedov phase after reverberation that is 
compatible with the radii of the CTA 1 PWN and SNR inside the parameter space that we have considered.
A better low and high-energy determination of the SED, in radio and GeV, in particular, would help distinguishing models better. CTA~1 thus remains as a difficult case, and a testbed for new models incorporating more complicated dynamics.

\subsection*{Acknowledgments}

We acknowledge support from the the grants  
AYA2015-71042-P
and SGR 2014--1073, and J. Li, E. de O\~na Wilhelmi and N. Rea for discussions.
We thank an anonymous referee for his/her remarks.

\bibliography{cta1_paper_v11}

\begin{thebibliography}{51}
\expandafter\ifx\csname natexlab\endcsname\relax\def\natexlab#1{#1}\fi

\bibitem[{{Abdo} {et~al}\mbox{.}(2008){Abdo}, {Ackermann}, {Atwood}, {Baldini},
  {Ballet}, {Barbiellini}, {Baring}, {Bastieri}, {Baughman}, {Bechtol},
  {Bellazzini}, {Berenji}, {Blandford}, {Bloom}, {Bogaert}, {Bonamente},
  {Borgland}, {Bregeon}, {Brez}, {Brigida}, {Bruel}, {Burnett}, {Caliandro},
  {Cameron}, {Caraveo}, {Carlson}, {Casandjian}, {Cecchi}, {Charles},
  {Chekhtman}, {Cheung}, {Chiang}, {Ciprini}, {Claus}, {Cohen-Tanugi},
  {Cominsky}, {Conrad}, {Cutini}, {Davis}, {Dermer}, {de Angelis}, {de Palma},
  {Digel}, {Dormody}, {do Couto e Silva}, {Drell}, {Dubois}, {Dumora},
  {Edmonds}, {Farnier}, {Focke}, {Fukazawa}, {Funk}, {Fusco}, {Gargano},
  {Gasparrini}, {Gehrels}, {Germani}, {Giebels}, {Giglietto}, {Giordano},
  {Glanzman}, {Godfrey}, {Grenier}, {Grondin}, {Grove}, {Guillemot}, {Guiriec},
  {Harding}, {Hartman}, {Hays}, {Hughes}, {J{\'o}hannesson}, {Johnson},
  {Johnson}, {Johnson}, {Johnson}, {Kamae}, {Kanai}, {Kanbach}, {Katagiri},
  {Kawai}, {Kerr}, {Kishishita}, {Kiziltan}, {Kn{\"o}dlseder}, {Kocian},
  {Komin}, {Kuehn}, {Kuss}, {Latronico}, {Lemoine-Goumard}, {Longo}, {Lonjou},
  {Loparco}, {Lott}, {Lovellette}, {Lubrano}, {Makeev}, {Marelli}, {Mazziotta},
  {McEnery}, {McGlynn}, {Meurer}, {Michelson}, {Mineo}, {Mitthumsiri},
  {Mizuno}, {Moiseev}, {Monte}, {Monzani}, {Morselli}, {Moskalenko}, {Murgia},
  {Nakamori}, {Nolan}, {Nuss}, {Ohno}, {Ohsugi}, {Okumura}, {Omodei},
  {Orlando}, {Ormes}, {Ozaki}, {Paneque}, {Panetta}, {Parent}, {Pelassa},
  {Pepe}, {Pesce-Rollins}, {Piano}, {Pieri}, {Piron}, {Porter}, {Rain{\`o}},
  {Rando}, {Ray}, {Razzano}, {Reimer}, {Reimer}, {Reposeur}, {Ritz},
  {Rochester}, {Rodriguez}, {Romani}, {Roth}, {Ryde}, {Sadrozinski}, {Sanchez},
  {Sander}, {Parkinson}, {Schalk}, {Sellerholm}, {Sgr{\`o}}, {Siskind},
  {Smith}, {Smith}, {Spandre}, {Spinelli}, {Starck}, {Strickman}, {Suson},
  {Tajima}, {Takahashi}, {Takahashi}, {Tanaka}, {Thayer}, {Thayer}, {Thompson},
  {Thorsett}, {Tibaldo}, {Torres}, {Tosti}, {Tramacere}, {Usher}, {Van Etten},
  {Vilchez}, {Vitale}, {Wang}, {Watters}, {Winer}, {Wood}, {Yasuda}, {Ylinen},
  \& {Ziegler}}]{abdo2008}
{Abdo} A.~A. {et~al.}, 2008, Science, 322, 1218

\bibitem[{{Abdo} {et~al}\mbox{.}(2012){Abdo}, {Wood}, {DeCesar}, {Gargano},
  {Giordano}, {Ray}, {Parent}, {Harding}, {Miller}, {Wood}, \&
  {Wolff}}]{abdo2012}
{Abdo} A.~A. {et~al.}, 2012, ApJ, 744, 146

\bibitem[{{Acero} {et~al}\mbox{.}(2013){Acero}, {Ackermann}, {Ajello},
  {Allafort}, {Baldini}, {Ballet}, {Barbiellini}, {Bastieri}, {Bechtol},
  {Bellazzini}, {Blandford}, {Bloom}, {Bonamente}, {Bottacini}, {Brandt},
  {Bregeon}, {Brigida}, {Bruel}, {Buehler}, {Buson}, {Caliandro}, {Cameron},
  {Caraveo}, {Cecchi}, {Charles}, {Chaves}, {Chekhtman}, {Chiang}, {Chiaro},
  {Ciprini}, {Claus}, {Cohen-Tanugi}, {Conrad}, {Cutini}, {Dalton},
  {D'Ammando}, {de Palma}, {Dermer}, {Di Venere}, {Silva}, {Drell},
  {Drlica-Wagner}, {Falletti}, {Favuzzi}, {Fegan}, {Ferrara}, {Focke},
  {Franckowiak}, {Fukazawa}, {Funk}, {Fusco}, {Gargano}, {Gasparrini},
  {Giglietto}, {Giordano}, {Giroletti}, {Glanzman}, {Godfrey}, {Gr{\'e}goire},
  {Grenier}, {Grondin}, {Grove}, {Guiriec}, {Hadasch}, {Hanabata}, {Harding},
  {Hayashida}, {Hayashi}, {Hays}, {Hewitt}, {Hill}, {Horan}, {Hou}, {Hughes},
  {Inoue}, {Jackson}, {Jogler}, {J{\'o}hannesson}, {Johnson}, {Kamae},
  {Kawano}, {Kerr}, {Kn{\"o}dlseder}, {Kuss}, {Lande}, {Larsson}, {Latronico},
  {Lemoine-Goumard}, {Longo}, {Loparco}, {Lovellette}, {Lubrano}, {Marelli},
  {Massaro}, {Mayer}, {Mazziotta}, {McEnery}, {Mehault}, {Michelson},
  {Mitthumsiri}, {Mizuno}, {Monte}, {Monzani}, {Morselli}, {Moskalenko},
  {Murgia}, {Nakamori}, {Nemmen}, {Nuss}, {Ohsugi}, {Okumura}, {Orienti},
  {Orlando}, {Ormes}, {Paneque}, {Panetta}, {Perkins}, {Pesce-Rollins},
  {Piron}, {Pivato}, {Porter}, {Rain{\`o}}, {Rando}, {Razzano}, {Reimer},
  {Reimer}, {Reposeur}, {Ritz}, {Roth}, {Rousseau}, {Saz Parkinson}, {Schulz},
  {Sgr{\`o}}, {Siskind}, {Smith}, {Spandre}, {Spinelli}, {Suson}, {Takahashi},
  {Takeuchi}, {Thayer}, {Thayer}, {Thompson}, {Tibaldo}, {Tibolla},
  {Tinivella}, {Torres}, {Tosti}, {Troja}, {Uchiyama}, {Vandenbroucke},
  {Vasileiou}, {Vianello}, {Vitale}, {Werner}, {Winer}, {Wood}, \&
  {Yang}}]{acero2013}
{Acero} F. {et~al.}, 2013, ApJ, 773, 77

\bibitem[{{Adriani} {et~al}\mbox{.}(2011){Adriani}, {Barbarino},
  {Bazilevskaya}, {Bellotti}, {Boezio}, {Bogomolov}, {Bonechi}, {Bongi},
  {Bonvicini}, {Borisov}, {Bottai}, {Bruno}, {Cafagna}, {Campana}, {Carbone},
  {Carlson}, {Casolino}, {Castellini}, {Consiglio}, {De Pascale}, {De Santis},
  {De Simone}, {Di Felice}, {Galper}, {Gillard}, {Grishantseva}, {Jerse},
  {Karelin}, {Koldashov}, {Krutkov}, {Kvashnin}, {Leonov}, {Malakhov},
  {Malvezzi}, {Marcelli}, {Mayorov}, {Menn}, {Mikhailov}, {Mocchiutti},
  {Monaco}, {Mori}, {Nikonov}, {Osteria}, {Palma}, {Papini}, {Pearce},
  {Picozza}, {Pizzolotto}, {Ricci}, {Ricciarini}, {Rossetto}, {Sarkar},
  {Simon}, {Sparvoli}, {Spillantini}, {Stozhkov}, {Vacchi}, {Vannuccini},
  {Vasilyev}, {Voronov}, {Yurkin}, {Wu}, {Zampa}, {Zampa}, \&
  {Zverev}}]{Adri2011}
{Adriani} O. {et~al.}, 2011, Science, 332, 69

\bibitem[{{Aharonian} {et~al}\mbox{.}(2008){Aharonian}, {Akhperjanian},
  {Bazer-Bachi}, {Behera}, {Beilicke}, {Benbow}, {Berge}, {Bernl{\"o}hr},
  {Boisson}, {Bolz}, {Borrel}, {Braun}, {Brion}, {Brown}, {B{\"u}hler},
  {Bulik}, {B{\"u}sching}, {Boutelier}, {Carrigan}, {Chadwick}, {Chounet},
  {Clapson}, {Coignet}, {Cornils}, {Costamante}, {Degrange}, {Dickinson},
  {Djannati-Ata{\"i}}, {Domainko}, {O'C.~Drury}, {Dubus}, {Dyks}, {Egberts},
  {Emmanoulopoulos}, {Espigat}, {Farnier}, {Feinstein}, {Fiasson},
  {F{\"o}rster}, {Fontaine}, {Fukui}, {Funk}, {Funk}, {F{\"u}{\ss}ling},
  {Gallant}, {Giebels}, {Glicenstein}, {Gl{\"u}ck}, {Goret}, {Hadjichristidis},
  {Hauser}, {Hauser}, {Heinzelmann}, {Henri}, {Hermann}, {Hinton}, {Hoffmann},
  {Hofmann}, {Holleran}, {Hoppe}, {Horns}, {Jacholkowska}, {de Jager},
  {Kendziorra}, {Kerschhaggl}, {Kh{\'e}lifi}, {Komin}, {Kosack}, {Lamanna},
  {Latham}, {Le Gallou}, {Lemi{\`e}re}, {Lemoine-Goumard}, {Lenain}, {Lohse},
  {Martin}, {Martineau-Huynh}, {Marcowith}, {Masterson}, {Maurin}, {McComb},
  {Moderski}, {Moriguchi}, {Moulin}, {de Naurois}, {Nedbal}, {Nolan}, {Olive},
  {Orford}, {Osborne}, {Ostrowski}, {Panter}, {Pedaletti}, {Pelletier},
  {Petrucci}, {Pita}, {P{\"u}hlhofer}, {Punch}, {Ranchon}, {Raubenheimer},
  {Raue}, {Rayner}, {Reimer}, {Renaud}, {Ripken}, {Rob}, {Rolland},
  {Rosier-Lees}, {Rowell}, {Rudak}, {Ruppel}, {Sahakian}, {Santangelo},
  {Saug{\'e}}, {Schlenker}, {Schlickeiser}, {Schr{\"o}der}, {Schwanke},
  {Schwarzburg}, {Schwemmer}, {Shalchi}, {Sol}, {Spangler}, {Stawarz},
  {Steenkamp}, {Stegmann}, {Superina}, {Takeuchi}, {Tam}, {Tavernet},
  {Terrier}, {van Eldik}, {Vasileiadis}, {Venter}, {Vialle}, {Vincent},
  {Vivier}, {V{\"o}lk}, {Volpe}, {Wagner}, \& {Ward}}]{Aha2008}
{Aharonian} F. {et~al.}, 2008, 481, 401

\bibitem[{{Aharonian}(1991)}]{aha_passive}
{Aharonian} F.~A., 1991, Ap\&SS, 180, 305

\bibitem[{{Aliu} {et~al}\mbox{.}(2013){Aliu}, {Archambault}, {Arlen}, {Aune},
  {Beilicke}, {Benbow}, {Bouvier}, {Buckley}, {Bugaev}, {Cesarini}, {Ciupik},
  {Collins-Hughes}, {Connolly}, {Cui}, {Dickherber}, {Duke}, {Dumm},
  {Dwarkadas}, {Errando}, {Falcone}, {Federici}, {Feng}, {Finley}, {Finnegan},
  {Fortson}, {Furniss}, {Galante}, {Gall}, {Gillanders}, {Godambe}, {Gotthelf},
  {Griffin}, {Grube}, {Gyuk}, {Hanna}, {Holder}, {Hughes}, {Humensky},
  {Kaaret}, {Kargaltsev}, {Karlsson}, {Khassen}, {Kieda}, {Krawczynski},
  {Krennrich}, {Lang}, {Lee}, {Madhavan}, {Maier}, {Majumdar}, {McArthur},
  {McCann}, {Moriarty}, {Mukherjee}, {Nelson}, {O'Faol{\'a}in de Bhr{\'o}ithe},
  {Ong}, {Orr}, {Otte}, {Park}, {Perkins}, {Pohl}, {Prokoph}, {Quinn}, {Ragan},
  {Reyes}, {Reynolds}, {Roache}, {Roberts}, {Saxon}, {Schroedter}, {Sembroski},
  {Slane}, {Smith}, {Staszak}, {Telezhinsky}, {Te{\v s}i{\'c}}, {Theiling},
  {Thibadeau}, {Tsurusaki}, {Tyler}, {Varlotta}, {Vassiliev}, {Vincent},
  {Vivier}, {Wakely}, {Weekes}, {Weinstein}, {Welsing}, {Williams}, \&
  {Zitzer}}]{aliu2013}
{Aliu} E. {et~al.}, 2013, ApJ, 764, 38

\bibitem[{{Bandiera}(1984)}]{bandiera1984}
{Bandiera} R., 1984, A\&A, 139, 368

\bibitem[{{Blondin}, {Chevalier} \& {Frierson}(2001){Blondin}, {Chevalier}, \&
  {Frierson}}]{blondin2001}
{Blondin} J.~M., {Chevalier} R.~A., {Frierson} D.~M., 2001, ApJ, 563, 806

\bibitem[{{Bolatto}, {Wolfire} \& {Leroy}(2013){Bolatto}, {Wolfire}, \&
  {Leroy}}]{xcoreview}
{Bolatto} A.~D., {Wolfire} M., {Leroy} A.~K., 2013, ARA\&A, 51, 207

\bibitem[{{Brazier} {et~al}\mbox{.}(1998){Brazier}, {Reimer}, {Kanbach}, \&
  {Carraminana}}]{brazier1998}
{Brazier} K.~T.~S., {Reimer} O., {Kanbach} G., {Carraminana} A., 1998, MNRAS,
  295, 819

\bibitem[{{Bucciantini}, {Arons} \& {Amato}(2011){Bucciantini}, {Arons}, \&
  {Amato}}]{bucciantini2011}
{Bucciantini} N., {Arons} J., {Amato} E., 2011, MNRAS, 410, 381

\bibitem[{{Caraveo} {et~al}\mbox{.}(2010){Caraveo}, {De Luca}, {Marelli},
  {Bignami}, {Ray}, {Saz Parkinson}, \& {Kanbach}}]{caraveo2010}
{Caraveo} P.~A., {De Luca} A., {Marelli} M., {Bignami} G.~F., {Ray} P.~S., {Saz
  Parkinson} P.~M., {Kanbach} G., 2010, ApJL, 725, L6

\bibitem[{{Chevalier}(2005)}]{chevalier2005}
{Chevalier} R.~A., 2005, ApJ, 619, 839

\bibitem[{{Chevalier} \& {Fransson}(1992)}]{chevalier1992}
{Chevalier} R.~A., {Fransson} C., 1992, ApJ, 395, 540

\bibitem[{{Condon} {et~al}\mbox{.}(1994){Condon}, {Broderick}, {Seielstad},
  {Douglas}, \& {Gregory}}]{gb6}
{Condon} J.~J., {Broderick} J.~J., {Seielstad} G.~A., {Douglas} K., {Gregory}
  P.~C., 1994, AJ, 107, 1829

\bibitem[{{Dame}(2011)}]{dame_mask}
{Dame} T.~M., 2011, ArXiv e-prints

\bibitem[{{Dame}, {Hartmann} \& {Thaddeus}(2001){Dame}, {Hartmann}, \&
  {Thaddeus}}]{newdame_co}
{Dame} T.~M., {Hartmann} D., {Thaddeus} P., 2001, ApJ, 547, 792

\bibitem[{{Dame} {et~al}\mbox{.}(1987){Dame}, {Ungerechts}, {Cohen}, {de Geus},
  {Grenier}, {May}, {Murphy}, {Nyman}, \& {Thaddeus}}]{dame_co}
{Dame} T.~M. {et~al.}, 1987, ApJ, 322, 706

\bibitem[{{de Cea del Pozo}, {Torres} \& {Rodriguez Marrero}(2009){de Cea del
  Pozo}, {Torres}, \& {Rodriguez Marrero}}]{decea}
{de Cea del Pozo} E., {Torres} D.~F., {Rodriguez Marrero} A.~Y., 2009, ApJ,
  698, 1054

\bibitem[{{de Jager} \& {Djannati-Ata{\"i}}(2009)}]{dejager2009}
{de Jager} O.~C., {Djannati-Ata{\"i}} A., 2009, in Astrophysics and Space
  Science Library, Vol. 357, Astrophysics and Space Science Library, {Becker}
  W., ed., p. 451

\bibitem[{{Dickey} \& {Lockman}(1990)}]{dickey1990}
{Dickey} J.~M., {Lockman} F.~J., 1990, ARA\&A, 28, 215

\bibitem[{{Domingo-Santamar{\'{\i}}a} \& {Torres}(2005)}]{dom2005}
{Domingo-Santamar{\'{\i}}a} E., {Torres} D.~F., 2005, A\&A, 444, 403

\bibitem[{{Gabici}(2008)}]{gabici_review}
{Gabici} S., 2008, ArXiv e-prints

\bibitem[{{Gaensler} \& {Slane}(2006)}]{gaensler2006}
{Gaensler} B.~M., {Slane} P.~O., 2006, ARA\&A, 44, 17

\bibitem[{{Gelfand}, {Slane} \& {Zhang}(2009){Gelfand}, {Slane}, \&
  {Zhang}}]{gelfand2009}
{Gelfand} J.~D., {Slane} P.~O., {Zhang} W., 2009, ApJ, 703, 2051

\bibitem[{{Giacani} {et~al}\mbox{.}(2014){Giacani}, {Rovero}, {Cillis},
  {Pichel}, \& {Dubner}}]{giacani2014}
{Giacani} E., {Rovero} A.~C., {Cillis} A., {Pichel} A., {Dubner} G., 2014,
  ArXiv e-prints

\bibitem[{{Halpern} {et~al}\mbox{.}(2004){Halpern}, {Gotthelf}, {Camilo},
  {Helfand}, \& {Ransom}}]{halpern2004}
{Halpern} J.~P., {Gotthelf} E.~V., {Camilo} F., {Helfand} D.~J., {Ransom}
  S.~M., 2004, ApJ, 612, 398

\bibitem[{{Harris} \& {Roberts}(1960)}]{harris1960}
{Harris} D.~E., {Roberts} J.~A., 1960, PASP, 72, 237

\bibitem[{{Lin} {et~al}\mbox{.}(2010){Lin}, {Huang}, {Takata}, {Hwang}, {Kong},
  \& {Hui}}]{lin2010}
{Lin} L.~C.~C., {Huang} R.~H.~H., {Takata} J., {Hwang} C.~Y., {Kong} A.~K.~H.,
  {Hui} C.~Y., 2010, ApJL, 725, L1

\bibitem[{{Lin} {et~al}\mbox{.}(2012){Lin}, {Takata}, {Kong}, {Hui}, {Enoto},
  {Chang}, {Huang}, {Liang}, {Shibata}, \& {Hwang}}]{lin2012}
{Lin} L.~C.~C. {et~al.}, 2012, MNRAS, 426, 2283

\bibitem[{{Martin} {et~al}\mbox{.}(2014){Martin}, {Torres}, {Cillis}, \& {de
  O{\~n}a Wilhelmi}}]{martin2014}
{Martin} J., {Torres} D.~F., {Cillis} A., {de O{\~n}a Wilhelmi} E., 2014,
  MNRAS, 443, 138

\bibitem[{{Mart{\'{\i}}n}, {Torres} \& {Rea}(2012){Mart{\'{\i}}n}, {Torres}, \&
  {Rea}}]{martin2012}
{Mart{\'{\i}}n} J., {Torres} D.~F., {Rea} N., 2012, MNRAS, 427, 415

\bibitem[{{Mattox} {et~al}\mbox{.}(1996){Mattox}, {Koh}, {Lamb}, {Macomb},
  {Prince}, \& {Ray}}]{mattox1996}
{Mattox} J.~R., {Koh} D.~T., {Lamb} R.~C., {Macomb} D.~J., {Prince} T.~A.,
  {Ray} P.~S., 1996, A\&AS, 120, C95

\bibitem[{{Mignani} {et~al}\mbox{.}(2013){Mignani}, {de Luca}, {Rea},
  {Shearer}, {Collins}, {Torres}, {Hadasch}, \& {Caliandro}}]{mignani2013}
{Mignani} R.~P., {de Luca} A., {Rea} N., {Shearer} A., {Collins} S., {Torres}
  D.~F., {Hadasch} D., {Caliandro} A., 2013, MNRAS, 430, 1354

\bibitem[{{Pedaletti} {et~al}\mbox{.}(2013){Pedaletti}, {Torres}, {Gabici}, {de
  O{\~n}a Wilhelmi}, {Mazin}, \& {Stamatescu}}]{gio}
{Pedaletti} G., {Torres} D.~F., {Gabici} S., {de O{\~n}a Wilhelmi} E., {Mazin}
  D., {Stamatescu} V., 2013, A\&A, 550, A123

\bibitem[{{Pineault} {et~al}\mbox{.}(1993){Pineault}, {Landecker}, {Madore}, \&
  {Gaumont-Guay}}]{pineault1993}
{Pineault} S., {Landecker} T.~L., {Madore} B., {Gaumont-Guay} S., 1993, AJ,
  105, 1060

\bibitem[{{Porter}, {Moskalenko} \& {Strong}(2006){Porter}, {Moskalenko}, \&
  {Strong}}]{porter2006}
{Porter} T.~A., {Moskalenko} I.~V., {Strong} A.~W., 2006, ApJL, 648, L29

\bibitem[{{Romero}, {Benaglia} \& {Torres}(1999){Romero}, {Benaglia}, \&
  {Torres}}]{romero99}
{Romero} G.~E., {Benaglia} P., {Torres} D.~F., 1999, A\&A, 348, 868

\bibitem[{{Seward}, {Schmidt} \& {Slane}(1995){Seward}, {Schmidt}, \&
  {Slane}}]{seward1995}
{Seward} F.~D., {Schmidt} B., {Slane} P., 1995, ApJ, 453, 284

\bibitem[{{Slane} {et~al}\mbox{.}(1997){Slane}, {Seward}, {Bandiera}, {Torii},
  \& {Tsunemi}}]{slane1997}
{Slane} P., {Seward} F.~D., {Bandiera} R., {Torii} K., {Tsunemi} H., 1997, ApJ,
  485, 221

\bibitem[{{Slane} {et~al}\mbox{.}(2004){Slane}, {Zimmerman}, {Hughes},
  {Seward}, {Gaensler}, \& {Clarke}}]{slane2004}
{Slane} P., {Zimmerman} E.~R., {Hughes} J.~P., {Seward} F.~D., {Gaensler}
  B.~M., {Clarke} M.~J., 2004, ApJ, 601, 1045

\bibitem[{{Torres} {et~al}\mbox{.}(2014){Torres}, {Cillis}, {Mart{\'{\i}}n}, \&
  {de O{\~n}a Wilhelmi}}]{torres2014}
{Torres} D.~F., {Cillis} A., {Mart{\'{\i}}n} J., {de O{\~n}a Wilhelmi} E.,
  2014, JHEAp, 1, 31

\bibitem[{{Torres}, {Cillis} \& {Mart{\'{\i}}n Rodriguez}(2013){Torres},
  {Cillis}, \& {Mart{\'{\i}}n Rodriguez}}]{torres2013a}
{Torres} D.~F., {Cillis} A.~N., {Mart{\'{\i}}n Rodriguez} J., 2013, ApJL, 763,
  L4

\bibitem[{{Torres} {et~al}\mbox{.}(2013){Torres}, {Mart{\'{\i}}n}, {de O{\~n}a
  Wilhelmi}, \& {Cillis}}]{torres2013b}
{Torres} D.~F., {Mart{\'{\i}}n} J., {de O{\~n}a Wilhelmi} E., {Cillis} A.,
  2013, MNRAS, 436, 3112

\bibitem[{{Torres}, {Rodriguez Marrero} \& {de Cea Del Pozo}(2008){Torres},
  {Rodriguez Marrero}, \& {de Cea Del Pozo}}]{ic443}
{Torres} D.~F., {Rodriguez Marrero} A.~Y., {de Cea Del Pozo} E., 2008, MNRAS,
  387, L59

\bibitem[{{Torres} {et~al}\mbox{.}(2003){Torres}, {Romero}, {Dame}, {Combi}, \&
  {Butt}}]{torres03}
{Torres} D.~F., {Romero} G.~E., {Dame} T.~M., {Combi} J.~A., {Butt} Y.~M.,
  2003, Phys. Rep., 382, 303

\bibitem[{{Truelove} \& {McKee}(1999)}]{truelove1999}
{Truelove} J.~K., {McKee} C.~F., 1999, ApJS, 120, 299

\bibitem[{{van der Swaluw} {et~al}\mbox{.}(2001){van der Swaluw}, {Achterberg},
  {Gallant}, \& {T{\'o}th}}]{Swaluw}
{van der Swaluw} E., {Achterberg} A., {Gallant} Y.~A., {T{\'o}th} G., 2001,
  A\&A, 380, 309

\bibitem[{{Zhang}, {Jiang} \& {Lin}(2009){Zhang}, {Jiang}, \&
  {Lin}}]{zhang2009}
{Zhang} L., {Jiang} Z.~J., {Lin} G.~F., 2009, ApJ, 699, 507

\bibitem[{{Ziegler} {et~al}\mbox{.}(2008){Ziegler}, {Baughman}, {Johnson}, \&
  {Atwood}}]{ziegler2008}
{Ziegler} M., {Baughman} B.~M., {Johnson} R.~P., {Atwood} W.~B., 2008, ApJ,
  680, 620

\end{thebibliography}

\label{lastpage}
\end{document}